\documentclass[final, 5p, twocolumn, authoryear]{elsarticle}

\usepackage{amsmath,amsfonts,amssymb,mathtools}
\usepackage{hyperref}
\usepackage[nameinlink,capitalise]{cleveref}
\usepackage{bm}
\usepackage{xcolor}
\usepackage{graphicx}
\usepackage{hyperref}
\usepackage{multirow}

\usepackage[scr]{rsfso}

\usepackage{cancel}

\journal{Physics of the Dark Universe}

\graphicspath{{./fig/}}

\usepackage{natbib}
\setcitestyle{numbers,square}
\usepackage[utf8]{inputenc}

\begin{document}
	\begin{frontmatter}
	
	\title{$\Lambda$CDM model against redshift-binned data: A mock analysis based on SNIa and Cosmic Chronometers}
	\author[label1]{Saeed Pourojaghi}
	\author[label1]{Mohammad Malekjani}
	\ead{malekjani@basu.ac.ir}
		
	\affiliation[label1]{
			organization={Department of Physics, Bu-Ali Sina University},
			city={Hamedan},
			postcode={65178, 016016},
			country={Iran}
		}

		\begin{abstract}
			Despite the broad successes of the flat $\Lambda$CDM model and its fitness to the various cosmological observations, it confronts challenges stemming from anomalies in the measurements of the Hubble constant ($H_0$) and the amplitude of matter fluctuations ($\sigma_8$). These inconsistencies have necessitated a reassessment of the model parameters, with a particular focus on their potential dependence on redshift.
			This study pioneers a new investigation to probe this redshift dependency by generating mock data simulated from observational data of Type Ia supernovae (SNIa) and cosmic chronometers (CC), thereby increasing the data density in this field. By sorting the data into high-redshift and low-redshift bins, we aim to refine the cosmological constraints on the parameters of the $\Lambda$CDM model and determine whether the noted dependence on redshift is due to a lack of high-redshift observational data or if they signify intrinsic issues within the model itself.
			Our approach employs the Markov Chain Monte Carlo (MCMC) algorithm to minimize the $\chi^2$ function, thus tightening the cosmological constraints. Our findings within the mock analysis reveal discrepancies between the values of $\Omega_{m0}$ and $H_0$ derived from the mock data bins with high redshift and low redshift, indicating the potential deviation of the standard $\Lambda$ CDM cosmology from the high-redshift SNIa and CC data. If this deviation proposes a new physics beyond the standard model, then with better quality future data tracking the new physics, these discrepancies will be statistically significant.

		\end{abstract}

		\begin{keyword}
			Cosmology, Dark energy, standard$\Lambda$CDM cosmology, Cosmological parameters
		\end{keyword}

	\end{frontmatter}

\section{Introduction}
The flat $\Lambda$CDM model stands as a remarkably simple yet profound framework that successfully accounts for the accelerated expansion of the universe. The consistency of this model with various observational data, including Type Ia supernovae (SNIa) \cite{Riess:1998cb,Perlmutter:1998np,Kowalski:2008ez}, cosmic microwave background (CMB) \cite{Komatsu2009,Jarosik:2010iu,Ade:2015rim}, weak gravitational lensing \cite{Benjamin:2007ys,Amendola:2007rr,Fu:2007qq}, baryon acoustic oscillations (BAO), and large-scale structure \cite{DESI:2024mwx, Tegmark:2003ud,Cole:2005sx,Eisenstein:2005su,Percival2010,Blake:2011en,Reid:2012sw}, is wonderful. In background level, the flat $\Lambda$CDM model simply explains the late-time history of the universe using just two parameters: the matter density parameter, $\Omega_{m0}$, and the Hubble constant, $H_0$, as described by the following Hubble parameter:

\begin{eqnarray}\label{eq1}
	H(z) = H_{0} \sqrt{\Omega_{m0}(1+z)^3 + (1-\Omega_{m0})}\;.
\end{eqnarray} 
	
Despite its successes, the $\Lambda$CDM model is not without some anomalies and challenges, which have prompted in detailed studies by cosmologists \cite{Weinberg:1988cp, Peebles:2002gy, Perivolaropoulos:2021jda, DiValentino:2021izs, Schoneberg:2021qvd, Abdalla:2022yfr, Akarsu:2024qiq}. A notable point of argument is the measured value of the Hubble constant from Cepheid variables at low redshift \cite{Riess:2021jrx}, which shows a significant tension -up to $5\sigma$- with the value inferred from CMB data at high redshift \cite{Planck:2018vyg}. Additionally, there is a discrepancy in the measurements of the weighted amplitude of matter fluctuations, denoted by $S_8$, with values from CMB experiments differing from those obtained through lensing surveys \cite{DES:2021wwk}, hinting at a potential inconsistency within the $\Lambda$CDM model.

Recent observations from DESI BAO \cite{DESI:2024mwx} and CMB anisotropic measurements \cite{Planck:2018vyg}, combined with various SNIa datasets (such as Pantheon+ \cite{Scolnic:2021amr}, Union 3 \cite{Rubin:2023ovl}, and DES 5YR SNIa \cite{DES:2024tys}), suggest potential deviations from the cosmological constant ($w_{\Lambda}=-1$). However, a study by the authors of \cite{DES:2024ywx}, using the absolute magnitude calibration of SNIa based on DESI BAO observations (instead of Cepheid), obtained a value of $H_0 = 67.19^{+0.66}_{-0.64} \; km/s/Mpc$ consistent with the Planck observations \cite{Planck:2018vyg}. It is important to note that this result assumes the Planck prior on the sound horizon parameter as a free parameter.

In recent years, a novel approach has been proposed to explore some anomalies within the $\Lambda$CDM model, utilizing observational data from low redshifts ($z < 3$). By binning the data alongside the redshift and constraining the free parameters of the cosmological model within each bin, it has been shown that the best-fit values of the cosmological parameters vary and exhibit a trend with effective redshift, a finding that defies the expectation of constancy in these parameters \cite{Colgain:2022rxy,Adil:2023jtu,Colgain:2023bge,Malekjani:2023ple,Colgain:2022tql,Colgain:2022nlb,Colgain:2024xqj,Colgain:2024ksa}.
 From a mathematical perspective, the present-time energy density $\rho_0$ acts as the integration constant derived from the continuity equation. This constant $\rho_0$ can be translated into a constant Hubble constant $H_0$ through the first Friedmann equation. Therefore, from an observational standpoint, it is expected that the $H_0$ value remains constant when determined using observational data at different redshifts. Furthermore, from observational point of view, the observational data from various redshifts should not indicate any variation in the constraints of $\Omega_{m0}$ and $H_0$ parameters. However, recent studies indicate a possibility of redshift-dependence in these parameters.

For example, \cite{Colgain:2022rxy} utilizing observational datasets from SNIa \cite{Pan-STARRS1:2017jku}, OHD \cite{Jimenez:2001gg,Seo:2003pu,SDSS:2005xqv}, and QSOs \cite{Lusso:2020pdb}, binning them into different redshift bins to put constraints on $H_0$ and $\Omega_{m0}$. They observed a redshift-trend of decreasing $H_0$ and increasing $\Omega_{m0}$ with effective redshift in the context of $\Lambda$CDM cosmology. Similarly, \cite{Malekjani:2023ple} demonstrated redshift evolution in $H_0$ and $\Omega_{m0}$ using the Pantheon+ sample within the framework of flat $\Lambda$CDM model. Their use of Bayesian analysis and profile distribution methods revealed an increasing statistical significance for this variation, suggesting the possibility of alternatives to the standard model or previously undetected systematic errors.
Furthermore, the increasement of $S_8$ with effective redshift within the $\Lambda$CDM framework has been shown by binning the growth rate data. A $2.8\sigma$ tension was found between the best-fit values of $S_8$ from high ($z > 1.1$) and low ($z < 1.1$) redshift data \cite{Adil:2023jtu}. The redshift evolution of the cosmological parameters within the standard $\Lambda$CDM model has been also explored utilizing the DESI BAO and DES SNIa data \cite{Colgain:2024xqj,Colgain:2024ksa}.

These studies collectively indicate that the $\Lambda$CDM model faces significant challenges. It is noteworthy that as redshift increases, the volume of observational data decreases markedly. For instance, the Pantheon+ sample \cite{Scolnic:2021amr} includes 1701 supernovae, with the vast majority detected at $z < 1$, and only 25 at $z > 1$, leading to increased uncertainty in the determination of cosmological parameters at higher redshifts. For example, the previous studies in binned analysis \cite{Colgain:2022rxy,Adil:2023jtu,Colgain:2023bge,Malekjani:2023ple,Colgain:2022tql,Colgain:2022nlb,Colgain:2024xqj,Colgain:2024ksa} suffer from low-density at high redshift bines. Therefore the observational constraints from high redshift bins have approximately high uncertainties leading to conclude the redshift-trend of cosmological parameters cautiously.
In this work, we aim to clear this issue by increasing the data density at higher redshifts through the simulation of mock data. We then equally divide the mock data into high-redshift ($z \geq 1$) and low-redshift ($z < 1$) bins, subsequently constraining the $\Lambda$CDM model's cosmological parameters for each bin. Our objective is to discern whether the redshift dependence of the parameters previously observed is dependent on the quantity of observational data. Should these redshift dependencies of the parameters be absent in our revised analysis, it would suggest a bias on data density. Conversely, their persistence would indicate potential defects in the cosmological standard $\Lambda$CDM model, proposing more further investigations into the fundamental nature of our universe.

The structure of this paper is organized as follows: Section (\ref{sect:mock}) details the process of our generating mock data. In Section (\ref{sec:result}), we apply observational constraints to the standard $\Lambda$CDM  model utilizing the generating mock data. Finally, in Section (\ref{conlusion}), we summarize our findings and conclude the study.

\section{Generating Mock data}\label{sect:mock}
In this section, we describe our methodology for generating mock data derived from real observational datasets. We produce two distinct samples of mock data: one set simulates the luminosity distance $d_L(z)$ using the observational data from Type Ia supernovae (SNIa) in the Pantheon+ catalog \cite{Scolnic:2021amr}, and the other set simulates the $H(z)$ data based on data from cosmic chronometers (CC) \cite{Colgain:2023bge}. Notice that CC data is a subset of OHD, which specifically uses the relative ages of passively evolving galaxies to estimate $\frac{dz}{dt}$ and thereby $H(z)$.
\subsection{Generate mock data based on Pantheon+ SNIa}\label{mock_SN}
\begin{figure} 
	\centering
	\includegraphics[width=9cm]{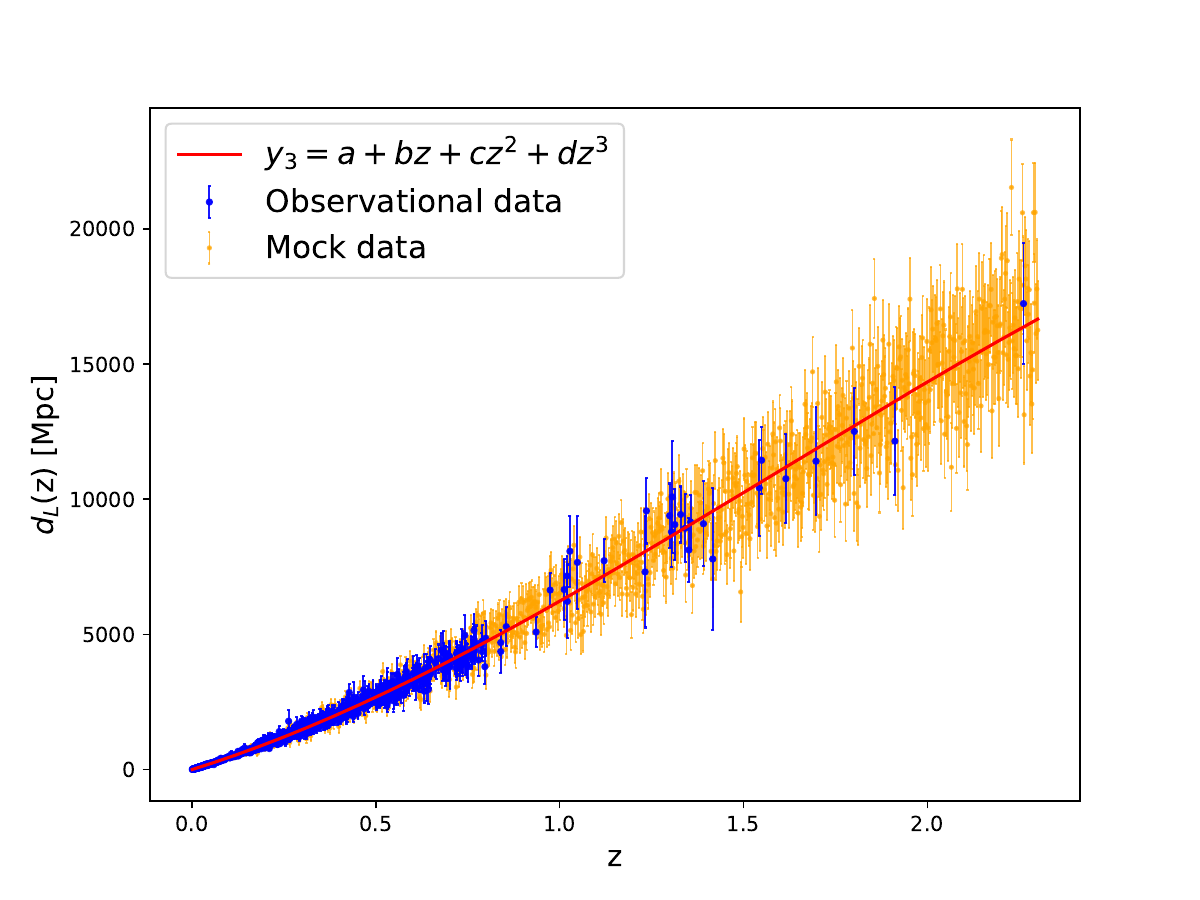}
	\caption{Comparative of Observational and Mock Data ($N=1000$) for $d_L$ Simulation.}
	\label{fig:dL}
\end{figure}

The Pantheon+ sample comprises 1701 individual supernovae within the redshift range $0 < z < 2.3$. 
The value of the distance modulus, $\mu$, and consequently the luminosity distance, $d_L$, of this data, are calculated by calibrating the absolute magnitude, $M_b$, with the SH0ES value.
To generate SNIa mock data, we first fit a third-order polynomial, $y_3(z) = a + b z + c z^2 + d z^3$, to the real observational $d_L$ data from the Pantheon+ catalog. We use the Cross-Validation analysis to determine how the third-order polynomial is chosen as the best polynomial fit to observations (see Appendix \ref{app:apx1}). Using the weighted least squares method, we determine the coefficients of the third-polynomial as $a = -1.8$, $b = 4224.7$, $c = 2537.1$, and $d = -531.7$. With $y_3(z)$ and its coefficients, we construct the mock $d_L$ data for SNIa.
We generated $N$ mock data points within the redshift range $0.001 \leq z \leq 2.3$. To achieve this, $y_3(z_i)$ is calculated for each specific redshift $z_i$. The corresponding mock luminosity distances, $d_L^{\text{mock}}(z_i)$, are then obtained by sampling from a Gaussian distribution, $\mathcal{N}(\mu, \sigma)$, with the mean value $\mu = y_3(z_i)$ and standard deviation $\sigma = \alpha$. Also, the error bar for $d_L^{\text{mock}}(z_i)$ is calculated as follows:
\begin{eqnarray}\label{eq3}
	d_L^{\text{mock}}(z_i) &=& \mathcal{N}(y_3(z_i),\; \alpha)\;, \nonumber \\
		\epsilon_{d_L^{\text{mock}}}(z_i) &=& \alpha \; d_L^{\text{mock}}(z_i)\;.
\end{eqnarray}

where $\alpha = 0.11$ is estimated as the average of $\frac{\epsilon_{d_L}}{d_L}$ from the observational SNIa data. In this context, $N$ represents the number of mock data points we have generated. We have produced 11 distinct sets of mock $d_L$ data, with $N$ ranging from 500 to 10,000. Figure \ref{fig:dL} displays 1000 mock $d_L$ data points generated using our method, alongside the observational data for comparison. For each dataset, we convert $d_L$ to distance modulus, $\mu$, using Eq. (\ref{eq4}). This allows us to constrain the cosmological parameters within the $\Lambda$CDM model using methods that will be detailed in the subsequent section.

\begin{eqnarray}\label{eq4}
	\mu_{\text{mock}} &=& 5\; log_{10}(d_L^{\text{mock}}) \; + \; 25\;, \nonumber \\
	\epsilon_{\mu_{\text{mock}}} &=& \frac{5}{log(10)} \; \dfrac{\epsilon_{d_L^{\text{mock}}}}{d_L^{\text{mock}}}\;.
\end{eqnarray}

\subsection{Generate mock data from CC observations}
Utilizing the methods outlined in the previous part, we are able to generate mock data based on observational data of cosmic chronometers. The CC data utilized in this study comprise 34 $H(z)$ data points within the redshift range $0.07 \leq z < 2$ \cite{2010JCAP...02..008S, 2012JCAP...08..006M, 2014RAA....14.1221Z, Moresco:2016mzx, Ratsimbazafy:2017vga, Borghi:2021zsr, Jiao:2022aep, Tomasetti:2023kek}.
As described in \ref{app:apx1}, based on Cross-Validation analysis, a second-order polynomial, $y_2(z) = a + b z + c z^2$, provides a good fit to the real observational CC data. Therefore, we use this polynomial to generate the CC mock data. By fitting $y_2(z)$ to the real observational CC data, we determine the coefficients of the polynomial to be $a = 63.6$, $b = 54.3$, and $c = 8.2$.
Using $y_2$, we calculate $N$ number of $y_2(z_i)$ for each specific $z_i$ in the redshift range $0.001 \leq z \leq 2$. Then, using a Gaussian distribution with mean value $y_2(z_i)$ and standard deviation $\beta$, we can create $H_{\text{mock}}(z_i)$. Where $\beta$ is the average of $\frac{\epsilon_H}{H}$ obtained form CC observational data.
\begin{eqnarray}\label{eq5}
	H_{\text{mock}}(z_i) &=& \mathcal{N}(y_2(z_i),\; \beta)\;, \nonumber \\
	\epsilon_{H_{\text{mock}}}(z_i) &=& \beta \; H_{\text{mock}}(z_i)\;.
\end{eqnarray}

\begin{figure} 
	\centering
	\includegraphics[width=9cm]{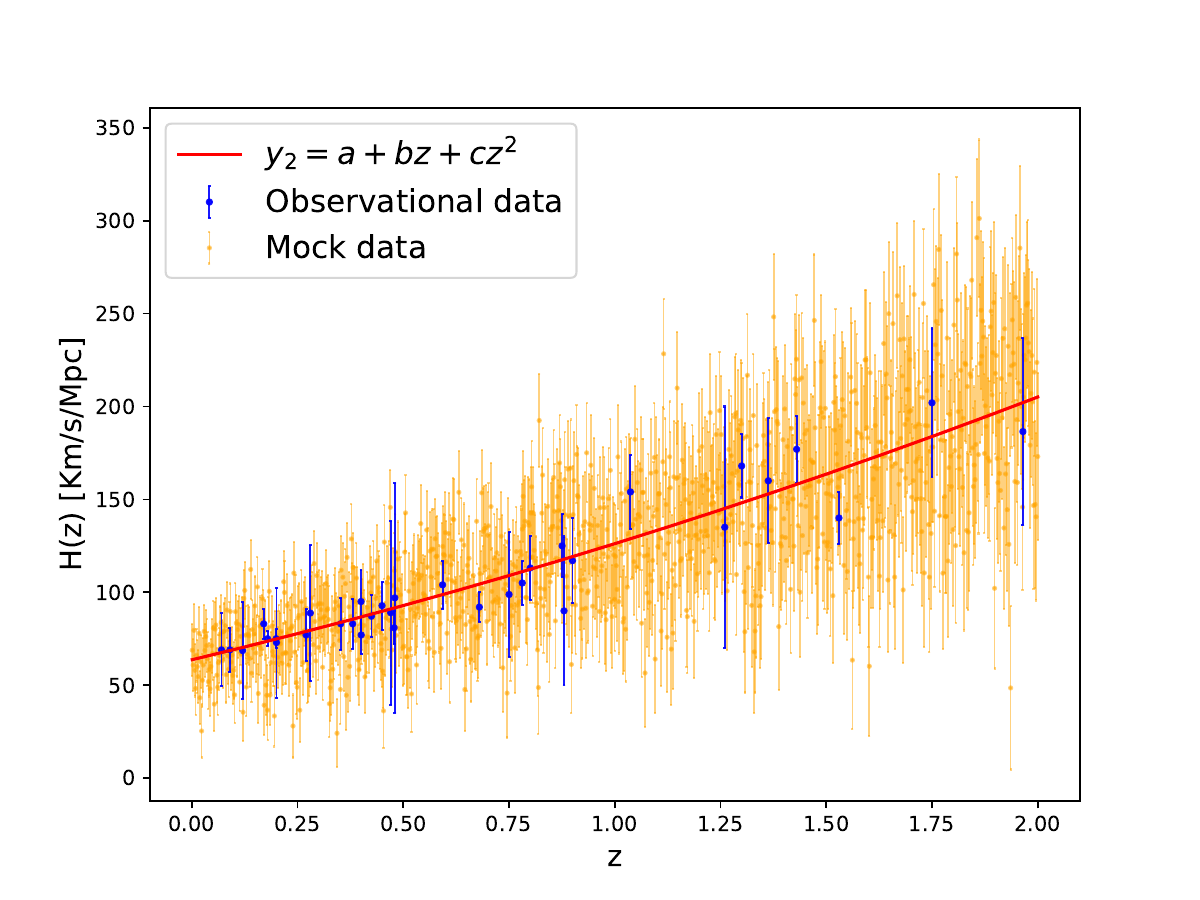}
	\caption{Same Figure \ref{fig:dL} But for $H(z)$ Simulation.}
	\label{fig:HH}
\end{figure}

Here we obtain $\beta=0.22$ from CC observations. It is worth noting that, as mentioned previously, $N$ is the number of mock data points that we generated. We created 11 sets of mock data based on CC, with $N$ ranging from 500 to 10,000. For instance, Figure \ref{fig:HH} shows a comparison of 1000 mock data points with real observational data.

Using these mock data in the next section, we explore potential deviations from Planck-$\Lambda$CDM cosmology (flat-$\Lambda$CDM model with $\Omega_{m0} \sim 0.3$) at higher redshifts, as previously observed in redshift-binned analyses using real observational data \cite{Colgain:2022rxy, Adil:2023jtu, Malekjani:2023ple, Colgain:2022tql, Colgain:2022nlb, Colgain:2024xqj, Colgain:2024ksa}. By performing polynomial fits to the SNIa and CC observational data, any deviations from the Planck-$\Lambda$CDM model at higher redshifts can be essentially encoded into our polynomial fits and subsequently into our mock data.
As pointed out in \cite{Colgain:2023bge}, the CC data at $z > 1$ prefer a horizontal $H(z)$ corresponding to $\Omega_{m0} = 0.0$. This indicates a significant deviation from Planck-$\Lambda$CDM cosmology using CC data at $z > 1$. Therefore, we expect to see these inherent deviations in our polynomial fit.
To investigate this, we fit-back the second-order polynomial to $N=10000$ mock CC data, divided into two bins: $5000$ data at $z < 1$ and $5000$ data at $z > 1$. We find the $\chi^2$ value for data at bin $z < 1$ as $5032$ and for data at bin $z > 1$ as $5217$. While all mock CC data fit a second-order polynomial (similar to what we see for the Planck-$\Lambda$CDM model in \cite{Colgain:2023bge}), we observe that our fitting deviates significantly ($\Delta \chi^2 = +185$) from the mock data at bin $z > 1$. Basically, we expect that this deviation leads to different results for constraining the cosmological parameters at low and high redshift bins.

\section{Cosmological Constraints in the $\Lambda$CDM Model through Mock Data}\label{sec:result}
In this section, we aim to constrain the cosmological parameters within a flat $\Lambda$CDM model utilizing mock data generated in the previous section. As previously mentioned, we produce 11 samples of mock data, varying in size from 500 to 10,000 data points. We have divided each sample into two equal bins: low-redshift data ($z < 1)$ and high-redshift data ($z \geq 1$), each subset $N/2$ data points. Additionally, we have compiled a full sample that includes all the generated Mock data. We then proceed to constrain the cosmological parameters $H_0$ and $\Omega_{m0}$ within the $\Lambda$CDM model for the full dataset, as well as separately for the high and low-redshift data in each sample. To achieve this, we employ the minimization of the $\chi^2$ function using the Markov Chain Monte Carlo (MCMC) algorithm.

\subsection{Numerical results for SNIa mock data}
As previously discussed, we can transform the luminosity distance into the distance modulus using Eq. \ref{eq4}. Once we have $\mu$, the $\chi^2$ function for this dataset can be expressed as follows:

\begin{eqnarray}\label{eq6}
	\chi^2_{SN} = \sum^{n}_{i=1}\dfrac{[\mu_{\text{mock}}(z_i)-\mu_{th}(z_i)]^2}{\epsilon_{\mu_{\text{mock},i}}^2}\;.
\end{eqnarray}

\begin{figure} 
	\centering
	\includegraphics[width=8.5cm]{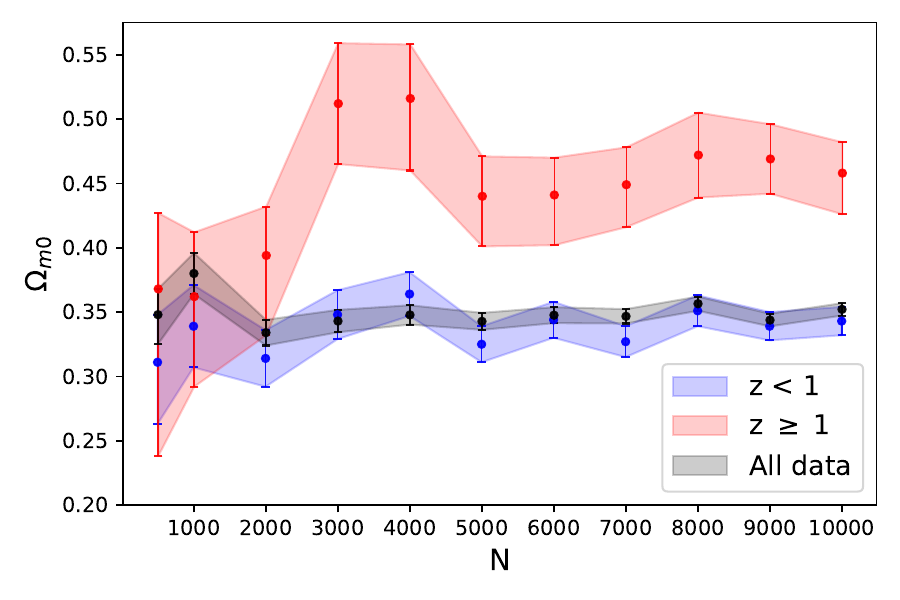}
	\caption{Schematic representation of best-fit $\Omega_{m0}$ with $1\sigma$ error, versus the variation of sample size of mock SNIa.}
	\label{fig:om_SN}
\end{figure}

\begin{figure} 
	\centering
	\includegraphics[width=8.5cm]{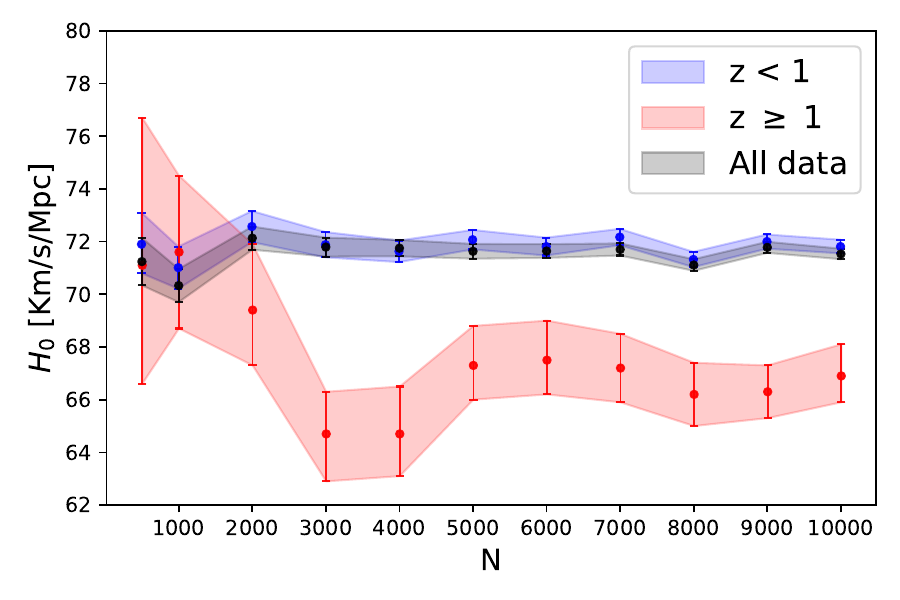}
	\caption{Same Figure \ref{fig:om_SN} but for $H_0$ Parameter.}
	\label{fig:H0_SN}
\end{figure}

\begin{figure} 
	\centering
	\includegraphics[width=9cm]{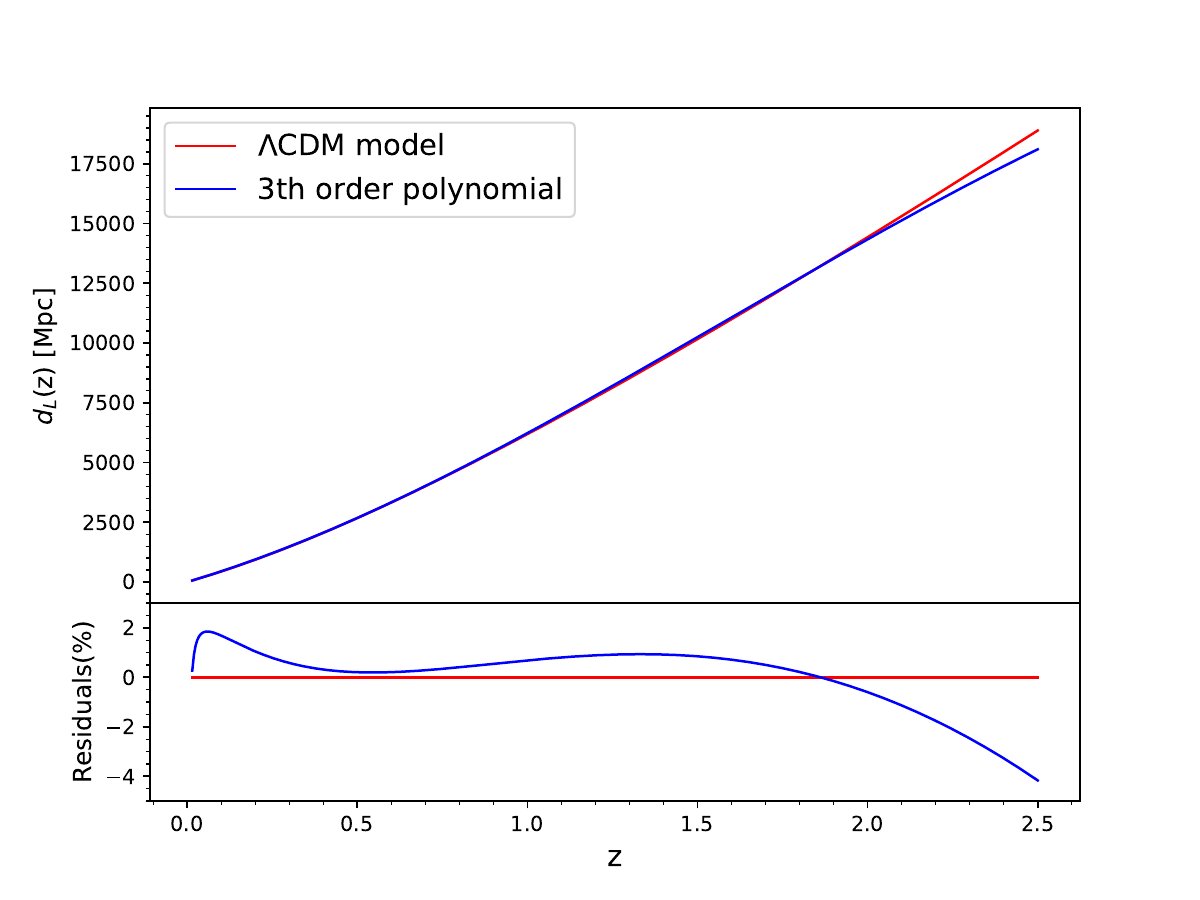}
	\caption{Comparing Luminosity Distance in the flat-$\Lambda$CDM Model and a Second-Order Polynomial.}
	\label{fig:residual}
\end{figure}

\begin{figure} 
	\centering
	\includegraphics[width=8cm]{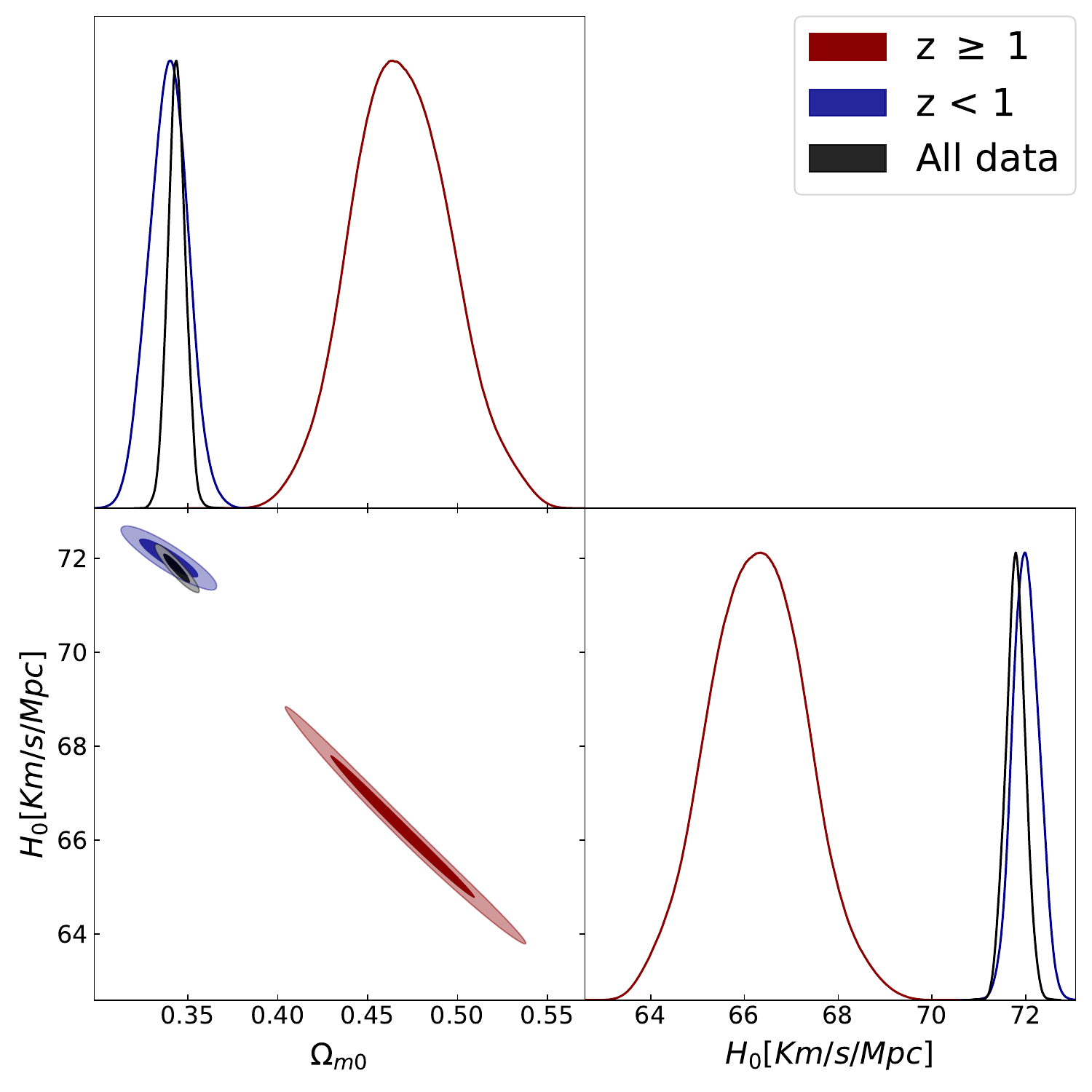}
	\caption{Confidence regions for $\Omega_{m0}-H_0$ in flat-$\Lambda$CDM model for a mock analysis with $N=9000$ SNIa data points.}
	\label{fig:9SN}
\end{figure}

Where $\mu_{\text{mock}}(z_i)$ represents the distance modulus of the mock data at redshift $z_i$, and $\epsilon_{\mu_{\text{mock},i}}$ denotes its error bar. Additionally, $\mu_{\text{th}}(z_i)$ is the theoretical value of the distance modulus at each specific redshift $z_i$, calculated as follows:

\begin{eqnarray}\label{eq7}
	\mu_{th}(z) = 5\; \log_{10}[(1+z) \; \frac{c}{H_0} \; \int_{0}^{z} \; \frac{dz'}{E(z')}] \; + \; 25\;.
\end{eqnarray}

where $c$ is the speed of light, and $E(z)=H/H_0$ is the dimensionless Hubble parameter, which is substituted from Eq. \ref{eq1} for the flat $\Lambda$CDM model.
Table \ref{res_SN} represents the numerical values of the best-fit cosmological parameters $\Omega_{m0}$ and $H_0$, along with their $1\sigma$ uncertainties obtained from our MCMC analysis for different mock SNIa datasets in different sub-samples at $z<1$ and $z \geq 1$. Our results reveal a discrepancy between the constraints from low and high redshift samples where an increase in the number of Mock data points correlates with a decrease in parameter uncertainties, as expected.
 For low-redshift ($z<1$) data, the best-fit values of $\Omega_{m0}$ and $H_0$ close to those obtained from the full dataset. In contrast, high-redshift ($z \geq 1$) data exhibit a higher $\Omega_{m0}$ and a lower $H_0$ compared to the low-redshift results, signifying a notable difference in the values of cosmological parameters. As depicted in Figures \ref{fig:om_SN} and \ref{fig:H0_SN}, the tension between low-redshift constraints and high-redshift constraints becomes more pronounced for $N \geq 3000$ dataset as the reduced uncertainties yield more precise results. In other words, when working with a small dataset, specifically where the number of data points is less than 2000, we find that the preferred values of cosmological parameters $\Omega_{m0}$ and $H_0$ at both high and low redshifts cannot be discriminated due to large uncertainties of our constraints. 
 It is worthwhile to note that the low number of mock data significantly affects the precision of the constraints on the cosmological parameters.
 This effect is particularly more pronounced in the case of high-redshift mock data sample, where the error bars of the data points are larger, leading to larger confidence levels of the constraints and less definitive conclusions. Our analysis shows that a larger numbers of data points causes to get more precised constraints on cosmological parameters throughout the entire range of redshifts. This may highlight the tension between the cosmological parameters derived from different redshift intervals. This predicted tension becomes more evident as the size of dataset grows, suggesting that there may be an intrinsic differences in the values of the cosmological parameters constrained from different redshift. This possible tension can be potentially assumed as a model breakdown of the $\Lambda$CDM cosmology. As mentioned before, when fitting a polynomial to observational data, the information is encoded into the polynomial. Any discrepancy between the observational data and the $\Lambda$CDM model can be observed as a difference between the polynomial fit and the model. In Figure \ref{fig:residual}, we show the difference between the luminosity distance of the best-fit flat-$\Lambda$CDM cosmology to observational SNIa data and the best-fit of the third-order polynomial fit to the observational SNIa data. As seen in the residual plane, the discrepancy between the two curves becomes more significant at higher redshifts. Consequently, when we generate mock data based on this polynomial, the difference form flat-$\Lambda$CDM model at high-redshifts encoded to the mock data. Hence, the tension between the measured cosmological parameters at high and low redshifts becomes apparent. Concretely, a tiny difference between the polynomial fit (reflecting the real observations) and the flat-$\Lambda$CDM model can be amplified to high statistical significance when using a larger number of data points. A similar pattern is observed for CC data, as discussed in Section \ref{sec:subCC}. As a typical example in Figure \ref{fig:9SN}, our analysis with a mock dataset comprising $N=9000$ data points reveals a notable and statistically significant tension between the constraints on the cosmological parameters obtained from the high-redshift and low-redshift data samples. Quantitatively speaking, we find $4.5\sigma$ deviation for $\Omega_{m0}$ and $5.5\sigma$ deviation for $H_0$ parameter derived from low and high redshift data samples. Such levels of observed tension suggest more puzzling issues in the instinct of the standard $\Lambda$CDM cosmological model and could potentially point to new physics or the need for a reevaluation of systematic uncertainties in the measurement processes of SNIa observations at different redshifts.
 
\begin{table*}
	\centering
	\caption{Estimated best-fit Values of Cosmological Parameters ($\Omega_{m0}$ and $H_0$) with $1\sigma$ confidence intervals in the flat-$\Lambda$CDM model obtained from different mock SNIa samples.}
	\begin{tabular}{c c c c c c c}
		\hline \hline
		\multirow{2}{*}{} & \multicolumn{2}{c}{$All \; data$} & \multicolumn{2}{c}{$low-z$} & \multicolumn{2}{c}{$high-z$}  \\
		\hline
		$n$  & $\Omega_{m0}$ & $H_0 [Km/s/Mpc]$ & $\Omega_{m0}$ & $H_0  [Km/s/Mpc]$ & $\Omega_{m0}$ & $H_0  [Km/s/Mpc]$ \\
		\hline
		$500$  & $0.348^{+0.020}_{-0.023}$ & $71.24\pm 0.90$ & $0.311^{+0.037}_{-0.048}$ & $71.9^{+1.2}_{-1.1}$ & $0.368^{+0.059}_{-0.13}$ & $71.1^{+5.6}_{-4.5}$ \\
		\hline
		$1000$  & $0.380\pm 0.016$ & $70.33\pm 0.63$ & $0.339\pm 0.032$ & $71.01\pm 0.80$ & $0.362^{+0.050}_{-0.070}$ & $71.6\pm 2.9$ \\
		\hline
		$2000$  & $0.334\pm 0.010$ & $72.13\pm 0.44$ & $0.314\pm 0.022$ & $72.57\pm 0.59$ & $0.394^{+0.038}_{-0.062}$ & $69.4^{+2.5}_{-2.1}$ \\
		\hline
		$3000$  & $0.3430\pm 0.0087$ & $71.79\pm 0.36$ & $0.348\pm 0.019$ & $71.88\pm 0.48$ & $0.512\pm 0.047$ & $64.7^{+1.6}_{-1.8}$ \\
		\hline
		$4000$  & $0.3478\pm 0.0075$ & $71.75\pm 0.31$ & $0.364\pm 0.017$ & $71.63\pm 0.41$ & $0.516^{+0.042}_{-0.056}$ & $64.7^{+1.8}_{-1.6}$ \\
		\hline
		$5000$  & $0.3428\pm 0.0066$ & $71.63\pm 0.28$ & $0.325\pm 0.014$ & $72.07\pm 0.37$ & $0.440^{+0.031}_{-0.039}$ & $67.3^{+1.5}_{-1.3}$ \\
		\hline
		$6000$  & $0.3476\pm 0.0062$ & $71.64\pm 0.26$ & $0.344\pm 0.014$ & $71.81\pm 0.34$ & $0.441^{+0.029}_{-0.039}$ & $67.5^{+1.5}_{-1.3}$ \\
		\hline
		$7000$  & $0.3467\pm 0.0056$ & $71.70\pm 0.23$ & $0.327\pm 0.012$ & $72.17\pm 0.31$ & $0.449^{+0.029}_{-0.033}$ & $67.2\pm 1.3$ \\
		\hline
		$8000$  & $0.3565\pm 0.0055$ & $71.11\pm 0.22$ & $0.351\pm 0.012$ & $71.32\pm 0.29$ & $0.472\pm 0.033$ & $66.2\pm 1.2$ \\
		\hline
		$9000$  & $0.3438\pm 0.0049$ & $71.78\pm 0.21$ & $0.339\pm 0.011$ & $72.00\pm 0.27$ & $0.469\pm 0.027$ & $66.3\pm 1.0$ \\
		\hline
		$10000$  & $0.3521\pm 0.0048$ & $71.53\pm 0.20$ & $0.343\pm 0.011$ & $71.81\pm 0.26$ & $0.458^{+0.024}_{-0.032}$ & $66.9^{+1.2}_{-1.0}$ \\     
		\hline \hline         
	\end{tabular}\label{res_SN}
\end{table*}

\subsection{Results for CC mock data}\label{sec:subCC}
In this part, we continue our analysis using mock CC data. The $\chi^2$ function for this data set can be written as follows:

\begin{eqnarray}\label{eq8}
	\chi^2_{CC} = \sum^{N}_{i=1}\dfrac{[H_{\text{mock}}(z_i)-H_{th}(z_i)]^2}{\epsilon_{H_{\text{mock},i}}^2}\;.
\end{eqnarray}

Where $H_{\text{mock}}(z_i)$ is the value of mock Hubble data in redshift $z_i$ and $\epsilon_{H_{\text{mock},i}}$ represents its associated error. Additionally, the theoretical value of the Hubble parameter at each specific redshift for a flat $\Lambda$CDM model is given by Eq. \ref{eq1}. 

Table \ref{res_CC} represents the best-fit values of $\Omega_{m0}$ and $H_0$ using various mock samples: low redshift data ($z < 1$), high redshift data ($z \geq 1$), and the entire dataset ($0<z \leq 2$). As previously mentioned, we utilize 11 datasets, each with varying number of mock data points from $N=500$ to $N=10000$.
As expected, the increasement in the number of data points leads to a reduction in the uncertainty of the measured parameters. Consequently, this enhances the precision of our constraints, allowing for more accurate determinations of cosmological parameters.

\begin{figure} 
	\centering
	\includegraphics[width=8.5cm]{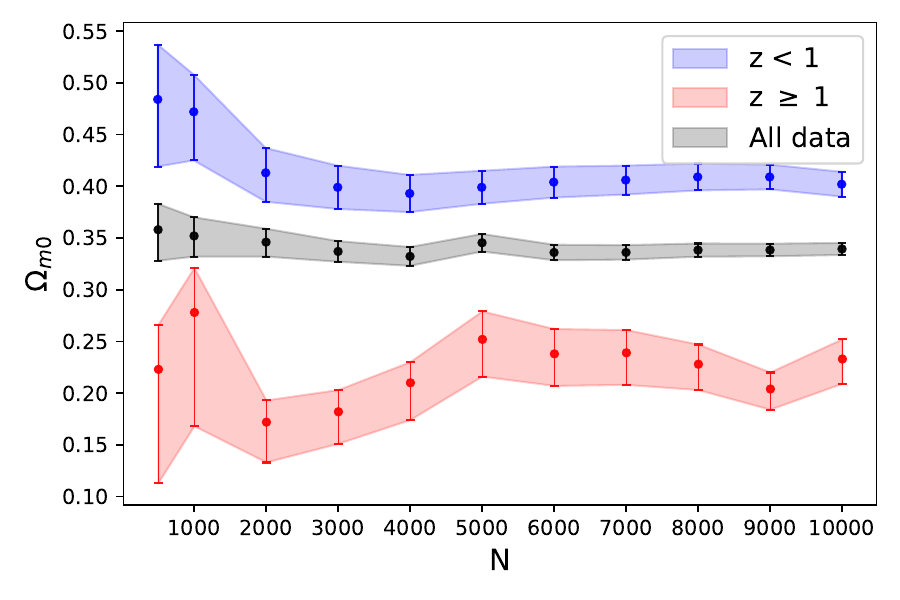}
	\caption{Same as Fig. \ref{fig:om_SN}, but for mock CC samples.}
	\label{fig:om_CC}
\end{figure}

\begin{figure} 
	\centering
	\includegraphics[width=8.5cm]{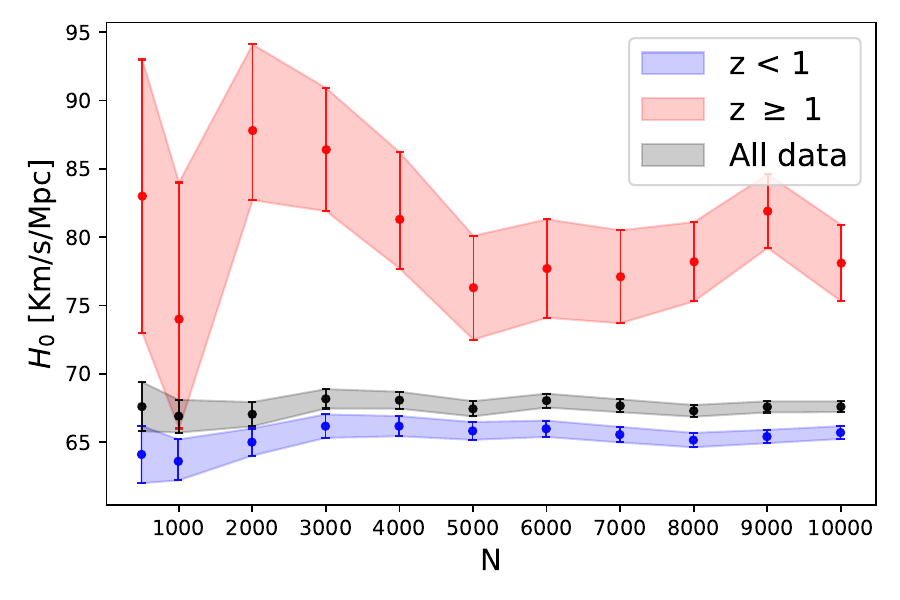}
	\caption{Same Figure \ref{fig:om_CC} But for $H_0$ Parameter.}
	\label{fig:H0_CC}
\end{figure}

\begin{figure} 
	\centering
	\includegraphics[width=8cm]{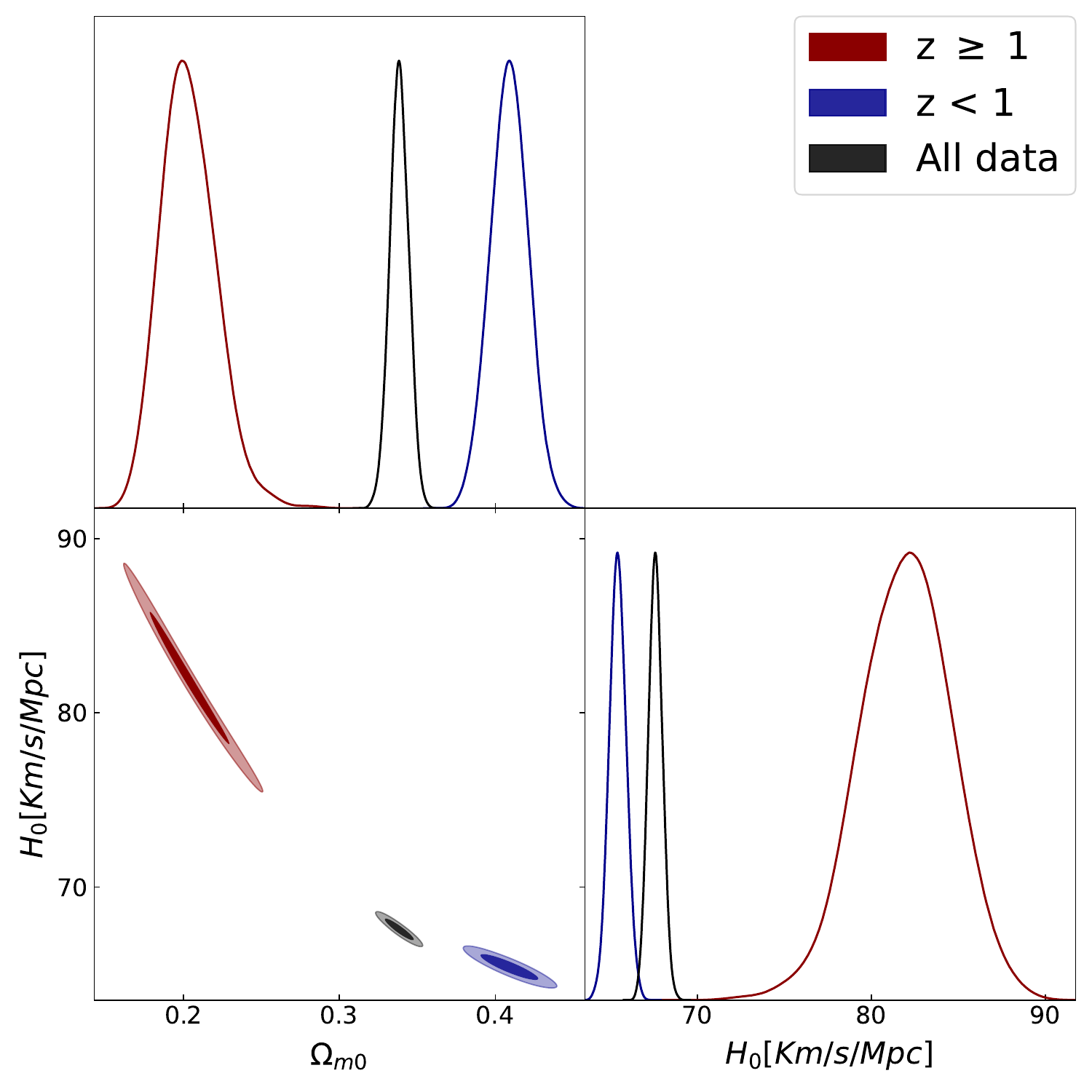}
	\caption{Same as Fig. \ref{fig:9SN}, but for mock CC data.}
	\label{fig:9CC}
\end{figure}

Upon analyzing the all data from all mock samples considered in Table \ref{res_CC}, we find that the derived values of $\Omega_{m0}$ and $H_0$ are in agreement with the Planck preferred cosmological parameter values, which report a Hubble constant $H_0 = 67.4 \pm 0.5$ $km/s/Mpc$ and a matter density parameter $\Omega_{m0} = 0.315 \pm 0.007$ \cite{Planck:2018vyg}. In a simple and quick glance, this consistency specially for mock samples with larger numbers of data indicates that the CC observations prefers the Planck experiments findings. However, exclusively utilizing high-redshift data, we observe a discernible trend: the value of the Hubble constant ($H_0$) tends to be higher, and the matter density parameter ($\Omega_{m0}$) correspondingly lower, compared to the values derived from the full dataset. Both predictions deviate from the Planck inferred values at high redshifts. Conversely, the values obtained from low-redshift mock data are almost consistent with those derived from the entire dataset. In other words, there is a noticeable trend in measuring cosmological parameters utilizing low-redshift to high-redshift CC observations in the context of flat $\Lambda$CDM model. The $H_0$ parameter exhibits an increasing trend, while $\Omega_{m0}$ shows a decreasing trend with respect to effective redshift.\\
As is clear in Figures \ref{fig:om_CC} and \ref{fig:H0_CC}, one can observe that the uncertainty of the constraints on the cosmological parameters reduces for larger samples of the mock data and the precision of the parameter estimation improves. This enhancement in precision, however, also brings into sharper focus the discrepancy between the values obtained from low-redshift and high-redshift data. Figure \ref{fig:9CC} further elucidates this point by presenting the confidence regions for $\Omega_{m0}$ and $H_0$ within the context of the flat $\Lambda$CDM model, utilizing a mock sample with number of $N=9000$ CC data. The corner plots distinctly show that the high-redshift and low-redshift data yield results that are significantly differ from each other. This difference manifests as a significant tension between the two sets of values, quantified as a $9.5\sigma$ discrepancy for $\Omega_{m0}$ and a $6\sigma$ discrepancy for $H_0$. Such substantial deviations suggest that there may be underlying systematic differences affecting the data at different redshifts, or they could potentially indicate a need for new physics beyond the standard flat $\Lambda$CDM cosmology to reconcile theories and observations.

\begin{table*}
	\centering
	\caption{Same Tab. \ref{res_SN} but for mock CC data.}
	\begin{tabular}{c c c c c c c}
		\hline \hline
		\multirow{2}{*}{} & \multicolumn{2}{c}{$All \; data$} & \multicolumn{2}{c}{$low-z$} & \multicolumn{2}{c}{$high-z$}  \\
		\hline
		$N$  & $\Omega_{m0}$ & $H_0 [Km/s/Mpc]$ & $\Omega_{m0}$ & $H_0 [Km/s/Mpc]$ & $\Omega_{m0}$ & $H_0 [Km/s/Mpc]$ \\
		\hline
		$500$  & $0.358^{+0.025}_{-0.030}$ & $67.6\pm 1.8$ & $0.484^{+0.053}_{-0.065}$ & $64.1\pm 2.1$ & $0.223^{+0.043}_{-0.11}$ & $83^{+10}_{-10}$ \\
		\hline
		$1000$ & $0.352^{+0.018}_{-0.020}$ & $66.9\pm 1.2$ & $0.472^{+0.036}_{-0.047}$ & $63.6^{+1.6}_{-1.4}$ & $0.278^{+0.043}_{-0.11}$ & $74^{+10}_{-8}$ \\
		\hline
		$2000$ & $0.346^{+0.013}_{-0.014}$ & $67.04\pm 0.89$ & $0.413^{+0.024}_{-0.028}$ & $65.0\pm 1.0$ & $0.172^{+0.021}_{-0.039}$ & $87.8^{+6.3}_{-5.1}$ \\
		\hline
		$3000$ & $0.337\pm 0.010$ & $68.17\pm 0.71$ & $0.399\pm 0.021$ & $66.17\pm 0.86$ & $0.182^{+0.021}_{-0.031}$ & $86.4\pm 4.5$ \\
		\hline
		$4000$ & $0.3322\pm 0.0091$ & $68.07\pm 0.62$ & $0.393\pm 0.018$ & $66.17\pm 0.74$ & $0.210^{+0.020}_{-0.036}$ & $81.3^{+4.9}_{-3.6}$ \\
		\hline
		$5000$ & $0.3454\pm 0.0085$ & $67.44\pm 0.56$ & $0.399\pm 0.016$ & $65.82\pm 0.65$ & $0.252^{+0.027}_{-0.036}$ & $76.3\pm 3.8$ \\
		\hline
		$6000$ & $0.3360\pm 0.0075$ & $68.04\pm 0.51$ & $0.404\pm 0.015$ & $65.98\pm 0.60$ & $0.238^{+0.024}_{-0.031}$ & $77.7\pm 3.6$ \\
		\hline
		$7000$ & $0.3361\pm 0.0070$ & $67.66\pm 0.47$ & $0.406\pm 0.014$ & $65.55\pm 0.57$ & $0.239^{+0.022}_{-0.031}$ & $77.1\pm 3.4$ \\
		\hline
		$8000$ & $0.3383\pm 0.0064$ & $67.29\pm 0.43$ & $0.409\pm 0.013$ & $65.15\pm 0.53$ & $0.228^{+0.019}_{-0.025}$ & $78.2\pm 2.9$ \\
		\hline
		$9000$ & $0.3384\pm 0.0061$ & $67.58\pm 0.41$ & $0.409\pm 0.012$ & $65.41\pm 0.49$ & $0.204^{+0.016}_{-0.020}$ & $81.9\pm 2.7$ \\
		\hline
		$10000$ & $0.3394\pm 0.0058$ & $67.59\pm 0.39$ & $0.402\pm 0.012$ & $65.70\pm 0.47$ & $0.233^{+0.019}_{-0.024}$ & $78.1\pm 2.8$ \\  
		\hline \hline         
	\end{tabular}\label{res_CC}
\end{table*}

\section{Conclusions} \label{conlusion}
In this study, we employed polynomial curve fitting techniques on the set of observational data including Type Ia Supernovae (SNIa) and Cosmic Chronometers (CC) to identify the most accurate representation of the data within the redshift range of $z < 2.5$. This process enabled us to generate a substantial big volume of mock data, which we then bin them into two distinct groups based on their redshift values: one for high redshift ($z \geq 1$) and another for low redshift ($z < 1$). Within these data binning, we put constraints on the free parameters of the flat $\Lambda$CDM model, $\Omega_{m0}$ and $H_0$, to study their redshift evolution reported by previous works \cite{Colgain:2022rxy, Adil:2023jtu, Colgain:2023bge, Malekjani:2023ple, Colgain:2022tql, Colgain:2022nlb,Colgain:2024xqj,Colgain:2024ksa} in a mock analysis with bigger data points but with the same errors.
Our findings revealed a notable discrepancy between the best-fit values of cosmological parameters when mock data were segregated into the aforementioned bins. This discrepancy suggests a potential trend in the $\Lambda$CDM cosmological parameters, either increasing or decreasing with redshift, which contradicts the expectation of their constancy as predicted by both mathematical framework and observational tests. This inconsistency points to a fundamental issue within the $\Lambda$CDM model itself.
It is worth noting that addressing the redshift evolution of cosmological parameters within $\Lambda$CDM model explored in \cite{Colgain:2022rxy, Adil:2023jtu, Colgain:2023bge, Malekjani:2023ple, Colgain:2022tql, Colgain:2022nlb,Colgain:2024xqj,Colgain:2024ksa} utilized the real observational data. 
However, their constraints on the cosmological parameters utilizing the higher redshift binned-data have bigger uncertainties due to low density of data. So the scarcity of real observational data at higher redshift bins typically decreases the accuracy of constraints of the cosmological parameters at those bins. So despite the observed trend of cosmological parameters with respect to redshift using the real observational data, one may connect this result to the low density of data at higher redshift. 
In this regard, our mock analysis indicates that the observed discrepancies between the cosmological constraints from low-redshift and high-redshift bins are not statistical anomalies and not due to paucity of data at high-redshift bins. Instead, they seem to be related from intrinsic issues with the standard $\Lambda$CDM cosmological model. By generating mock data, we were able to enhance the data density at higher redshift zones, thereby supporting the conclusion that the tensions between predictions of low-redshift and high-redshift bins arise within the framework of $\Lambda$CDM cosmology. This finding emphasizes the need for a re-evaluation of the $\Lambda$CDM model or the consideration of alternative models that can reduce or alleviate the redshift-dependent behavior of the cosmological parameters.

\section{Acknowledgment}
We would like to extend our gratitude to the anonymous referees for their invaluable feedback and constructive comments, which have significantly enhanced the quality of this work. The authors also would like to express their gratitude to E. \'O. Colg\'ain and M. M. Sheikh-Jabbari for their valuable discussions. This work is based upon research funded by the Iran National Science Foundation (INSF) under project No. 4024802. 

\bibliographystyle{elsarticle-num}
\bibliography{ref}

\begin{thebibliography}{51}
\providecommand{\natexlab}[1]{#1}
\providecommand{\url}[1]{\texttt{#1}}
\expandafter\ifx\csname urlstyle\endcsname\relax
  \providecommand{\doi}[1]{doi: #1}\else
  \providecommand{\doi}{doi: \begingroup \urlstyle{rm}\Url}\fi

\bibitem[Riess et~al.(1998)]{Riess:1998cb}
Adam~G. Riess et~al.
\newblock {Observational evidence from supernovae for an accelerating universe
  and a cosmological constant}.
\newblock \emph{Astron. J.}, 116:\penalty0 1009--1038, 1998.

\bibitem[Perlmutter et~al.(1999)]{Perlmutter:1998np}
S.~Perlmutter et~al.
\newblock {Measurements of $\Omega$ and $\Lambda$ from 42 high redshift
  supernovae}.
\newblock \emph{Astrophys. J.}, 517:\penalty0 565--586, 1999.

\bibitem[Kowalski et~al.(2008)]{Kowalski:2008ez}
M.~Kowalski et~al.
\newblock {Improved Cosmological Constraints from New, Old and Combined
  Supernova Datasets}.
\newblock \emph{Astrophys. J.}, 686:\penalty0 749--778, 2008.

\bibitem[Komatsu et~al.(2009)Komatsu, Dunkley, Nolta, and et~al.]{Komatsu2009}
E.~Komatsu, J.~Dunkley, M.~R. Nolta, and et~al.
\newblock \emph{ApJS}, 180:\penalty0 330, 2009.

\bibitem[Jarosik et~al.(2011)Jarosik, Bennett, Dunkley, Gold, Greason, Halpern,
  Hill, Hinshaw, Kogut, Komatsu, and et~al.]{Jarosik:2010iu}
N.~Jarosik, C.~L. Bennett, J.~Dunkley, B.~Gold, M.~R. Greason, M.~Halpern,
  R.~S. Hill, G.~Hinshaw, A.~Kogut, E.~Komatsu, and et~al.
\newblock \emph{apjs}, 192:\penalty0 14, 2011.

\bibitem[Ade et~al.(2016)]{Ade:2015rim}
P.~A.~R. Ade et~al.
\newblock {Planck 2015 results. XIV. Dark energy and modified gravity}.
\newblock \emph{Astron. Astrophys.}, 594:\penalty0 A14, 2016.

\bibitem[Benjamin et~al.(2007)Benjamin, Heymans, Semboloni, Van~Waerbeke,
  Hoekstra, Erben, Gladders, Hetterscheidt, Mellier, and Yee]{Benjamin:2007ys}
Jonathan Benjamin, Catherine Heymans, Elisabetta Semboloni, Ludovic
  Van~Waerbeke, Henk Hoekstra, Thomas Erben, Michael~D. Gladders, Marco
  Hetterscheidt, Yannick Mellier, and H.~K.~C. Yee.
\newblock {Cosmological Constraints From the 100 Square Degree Weak Lensing
  Survey}.
\newblock \emph{Mon. Not. Roy. Astron. Soc.}, 381:\penalty0 702--712, 2007.

\bibitem[Amendola et~al.(2008)Amendola, Kunz, and Sapone]{Amendola:2007rr}
Luca Amendola, Martin Kunz, and Domenico Sapone.
\newblock {Measuring the dark side (with weak lensing)}.
\newblock \emph{JCAP}, 0804:\penalty0 013, 2008.

\bibitem[Fu et~al.(2008)]{Fu:2007qq}
L.~Fu et~al.
\newblock {Very weak lensing in the CFHTLS Wide: Cosmology from cosmic shear in
  the linear regime}.
\newblock \emph{Astron. Astrophys.}, 479:\penalty0 9--25, 2008.

\bibitem[Adame et~al.(2024)]{DESI:2024mwx}
A.~G. Adame et~al.
\newblock {DESI 2024 VI: Cosmological Constraints from the Measurements of
  Baryon Acoustic Oscillations}.
\newblock 4 2024.

\bibitem[Tegmark et~al.(2004)]{Tegmark:2003ud}
Max Tegmark et~al.
\newblock {Cosmological parameters from SDSS and WMAP}.
\newblock \emph{Phys. Rev. D}, 69:\penalty0 103501, 2004.

\bibitem[Cole et~al.(2005)]{Cole:2005sx}
Shaun Cole et~al.
\newblock {The 2dF Galaxy Redshift Survey: Power-spectrum analysis of the final
  dataset and cosmological implications}.
\newblock \emph{MNRAS}, 362:\penalty0 505--534, 2005.

\bibitem[Eisenstein et~al.(2005{\natexlab{a}})]{Eisenstein:2005su}
Daniel~J. Eisenstein et~al.
\newblock {Detection of the baryon acoustic peak in the large-scale correlation
  function of SDSS luminous red galaxies}.
\newblock \emph{ApJ}, 633:\penalty0 560--574, 2005{\natexlab{a}}.

\bibitem[Percival et~al.(2010)Percival, Reid, Eisenstein, and
  et~al.]{Percival2010}
W.~J. Percival, B.~A. Reid, D.~J. Eisenstein, and et~al.
\newblock \emph{\mnras}, 401:\penalty0 2148, 2010.

\bibitem[Blake et~al.(2011)Blake, Kazin, Beutler, Davis, Parkinson,
  et~al.]{Blake:2011en}
Chris Blake, Eyal Kazin, Florian Beutler, Tamara Davis, David Parkinson, et~al.
\newblock {The WiggleZ Dark Energy Survey: mapping the distance-redshift
  relation with baryon acoustic oscillations}.
\newblock \emph{MNRAS}, 418:\penalty0 1707--1724, 2011.

\bibitem[Reid et~al.(2012)Reid, Samushia, White, Percival, Manera,
  et~al.]{Reid:2012sw}
Beth~A. Reid, Lado Samushia, Martin White, Will~J. Percival, Marc Manera,
  et~al.
\newblock {The clustering of galaxies in the SDSS-III Baryon Oscillation
  Spectroscopic Survey: measurements of the growth of structure and expansion
  rate at z=0.57 from anisotropic clustering}.
\newblock \emph{MNRAS}, 426:\penalty0 2719, 2012.

\bibitem[Weinberg(1989)]{Weinberg:1988cp}
Steven Weinberg.
\newblock {The Cosmological Constant Problem}.
\newblock \emph{Rev. Mod. Phys.}, 61:\penalty0 1--23, 1989.
\newblock \doi{10.1103/RevModPhys.61.1}.

\bibitem[Peebles and Ratra(2003)]{Peebles:2002gy}
P.~J.~E. Peebles and Bharat Ratra.
\newblock {The Cosmological Constant and Dark Energy}.
\newblock \emph{Rev. Mod. Phys.}, 75:\penalty0 559--606, 2003.
\newblock \doi{10.1103/RevModPhys.75.559}.

\bibitem[Perivolaropoulos and Skara(2022)]{Perivolaropoulos:2021jda}
Leandros Perivolaropoulos and Foteini Skara.
\newblock {Challenges for \ensuremath{\Lambda}CDM: An update}.
\newblock \emph{New Astron. Rev.}, 95:\penalty0 101659, 2022.
\newblock \doi{10.1016/j.newar.2022.101659}.

\bibitem[Di~Valentino et~al.(2021)Di~Valentino, Mena, Pan, Visinelli, Yang,
  Melchiorri, Mota, Riess, and Silk]{DiValentino:2021izs}
Eleonora Di~Valentino, Olga Mena, Supriya Pan, Luca Visinelli, Weiqiang Yang,
  Alessandro Melchiorri, David~F. Mota, Adam~G. Riess, and Joseph Silk.
\newblock {In the realm of the Hubble tension\textemdash{}a review of
  solutions}.
\newblock \emph{Class. Quant. Grav.}, 38\penalty0 (15):\penalty0 153001, 2021.
\newblock \doi{10.1088/1361-6382/ac086d}.

\bibitem[Sch\"oneberg et~al.(2022)Sch\"oneberg, Franco~Abell\'an,
  P\'erez~S\'anchez, Witte, Poulin, and Lesgourgues]{Schoneberg:2021qvd}
Nils Sch\"oneberg, Guillermo Franco~Abell\'an, Andrea P\'erez~S\'anchez,
  Samuel~J. Witte, Vivian Poulin, and Julien Lesgourgues.
\newblock {The H0 Olympics: A fair ranking of proposed models}.
\newblock \emph{Phys. Rept.}, 984:\penalty0 1--55, 2022.
\newblock \doi{10.1016/j.physrep.2022.07.001}.

\bibitem[Abdalla et~al.(2022)]{Abdalla:2022yfr}
Elcio Abdalla et~al.
\newblock {Cosmology intertwined: A review of the particle physics,
  astrophysics, and cosmology associated with the cosmological tensions and
  anomalies}.
\newblock \emph{JHEAp}, 34:\penalty0 49--211, 2022.
\newblock \doi{10.1016/j.jheap.2022.04.002}.

\bibitem[Akarsu et~al.(2024)Akarsu, Colg\'ain, Sen, and
  Sheikh-Jabbari]{Akarsu:2024qiq}
\"Ozg\"ur Akarsu, Eoin~\'O. Colg\'ain, Anjan~A. Sen, and M.~M. Sheikh-Jabbari.
\newblock {$\Lambda$CDM Tensions: Localising Missing Physics through
  Consistency Checks}.
\newblock 2 2024.

\bibitem[Riess et~al.(2022)]{Riess:2021jrx}
Adam~G. Riess et~al.
\newblock {A Comprehensive Measurement of the Local Value of the Hubble
  Constant with 1 km/s/Mpc Uncertainty from the Hubble Space Telescope and the
  SH0ES Team}.
\newblock \emph{Astrophys. J. Lett.}, 934\penalty0 (1):\penalty0 L7, 2022.
\newblock \doi{10.3847/2041-8213/ac5c5b}.

\bibitem[Aghanim et~al.(2020)]{Planck:2018vyg}
N.~Aghanim et~al.
\newblock {Planck 2018 results. VI. Cosmological parameters}.
\newblock \emph{Astron. Astrophys.}, 641:\penalty0 A6, 2020.
\newblock \doi{10.1051/0004-6361/201833910}.
\newblock [Erratum: Astron.Astrophys. 652, C4 (2021)].

\bibitem[Abbott et~al.(2022)]{DES:2021wwk}
T.~M.~C. Abbott et~al.
\newblock {Dark Energy Survey Year 3 results: Cosmological constraints from
  galaxy clustering and weak lensing}.
\newblock \emph{Phys. Rev. D}, 105\penalty0 (2):\penalty0 023520, 2022.
\newblock \doi{10.1103/PhysRevD.105.023520}.

\bibitem[Scolnic et~al.(2022)]{Scolnic:2021amr}
Dan Scolnic et~al.
\newblock {The Pantheon+ Analysis: The Full Data Set and Light-curve Release}.
\newblock \emph{Astrophys. J.}, 938\penalty0 (2):\penalty0 113, 2022.
\newblock \doi{10.3847/1538-4357/ac8b7a}.

\bibitem[Rubin et~al.(2023)]{Rubin:2023ovl}
David Rubin et~al.
\newblock {Union Through UNITY: Cosmology with 2,000 SNe Using a Unified
  Bayesian Framework}.
\newblock 11 2023.

\bibitem[Abbott et~al.(2024)]{DES:2024tys}
T.~M.~C. Abbott et~al.
\newblock {The Dark Energy Survey: Cosmology Results With \textasciitilde{}1500
  New High-redshift Type Ia Supernovae Using The Full 5-year Dataset}.
\newblock 1 2024.

\bibitem[Camilleri et~al.(2024)]{DES:2024ywx}
R.~Camilleri et~al.
\newblock {The Dark Energy Survey Supernova Program: An updated measurement of
  the Hubble constant using the Inverse Distance Ladder}.
\newblock 6 2024.

\bibitem[Colg\'ain et~al.(2024{\natexlab{a}})Colg\'ain, Sheikh-Jabbari,
  Solomon, Dainotti, and Stojkovic]{Colgain:2022rxy}
Eoin~\'O. Colg\'ain, M.~M. Sheikh-Jabbari, Rance Solomon, Maria~G. Dainotti,
  and Dejan Stojkovic.
\newblock {Putting flat \ensuremath{\Lambda}CDM in the (Redshift) bin}.
\newblock \emph{Phys. Dark Univ.}, 44:\penalty0 101464, 2024{\natexlab{a}}.
\newblock \doi{10.1016/j.dark.2024.101464}.

\bibitem[Adil et~al.(2023)Adil, Akarsu, Malekjani, Colg\'ain, Pourojaghi, Sen,
  and Sheikh-Jabbari]{Adil:2023jtu}
Shahnawaz~A. Adil, \"Ozg\"ur Akarsu, Mohammad Malekjani, Eoin~\'O. Colg\'ain,
  Saeed Pourojaghi, Anjan~A. Sen, and M.~M. Sheikh-Jabbari.
\newblock {S8 increases with effective redshift in \ensuremath{\Lambda}CDM
  cosmology}.
\newblock \emph{Mon. Not. Roy. Astron. Soc.}, 528\penalty0 (1):\penalty0
  L20--L26, 2023.
\newblock \doi{10.1093/mnrasl/slad165}.

\bibitem[Colg\'ain et~al.(2023{\natexlab{a}})Colg\'ain, Pourojaghi,
  Sheikh-Jabbari, and Sherwin]{Colgain:2023bge}
Eoin~\'O. Colg\'ain, Saeed Pourojaghi, M.~M. Sheikh-Jabbari, and Darragh
  Sherwin.
\newblock {MCMC Marginalisation Bias and $\Lambda$CDM tensions}.
\newblock 7 2023{\natexlab{a}}.

\bibitem[Malekjani et~al.(2024)Malekjani, Conville, Colg\'ain, Pourojaghi, and
  Sheikh-Jabbari]{Malekjani:2023ple}
Mohammad Malekjani, Ruair\'\i{}~Mc Conville, Eoin~\'O. Colg\'ain, Saeed
  Pourojaghi, and M.~M. Sheikh-Jabbari.
\newblock {On redshift evolution and negative dark energy density in Pantheon +
  Supernovae}.
\newblock \emph{Eur. Phys. J. C}, 84\penalty0 (3):\penalty0 317, 2024.
\newblock \doi{10.1140/epjc/s10052-024-12667-z}.

\bibitem[Colg\'ain et~al.(2023{\natexlab{b}})Colg\'ain, Sheikh-Jabbari, and
  Solomon]{Colgain:2022tql}
Eoin~\'O. Colg\'ain, M.~M. Sheikh-Jabbari, and Rance Solomon.
\newblock {High redshift \ensuremath{\Lambda}CDM cosmology: To bin or not to
  bin?}
\newblock \emph{Phys. Dark Univ.}, 40:\penalty0 101216, 2023{\natexlab{b}}.
\newblock \doi{10.1016/j.dark.2023.101216}.

\bibitem[Colg\'ain et~al.(2022)Colg\'ain, Sheikh-Jabbari, Solomon, Bargiacchi,
  Capozziello, Dainotti, and Stojkovic]{Colgain:2022nlb}
Eoin~\'O. Colg\'ain, M.~M. Sheikh-Jabbari, Rance Solomon, Giada Bargiacchi,
  Salvatore Capozziello, Maria~Giovanna Dainotti, and Dejan Stojkovic.
\newblock {Revealing intrinsic flat \ensuremath{\Lambda}CDM biases with
  standardizable candles}.
\newblock \emph{Phys. Rev. D}, 106\penalty0 (4):\penalty0 L041301, 2022.
\newblock \doi{10.1103/PhysRevD.106.L041301}.

\bibitem[Colg\'ain et~al.(2024{\natexlab{b}})Colg\'ain, Dainotti, Capozziello,
  Pourojaghi, Sheikh-Jabbari, and Stojkovic]{Colgain:2024xqj}
Eoin~\'O. Colg\'ain, Maria~Giovanna Dainotti, Salvatore Capozziello, Saeed
  Pourojaghi, M.~M. Sheikh-Jabbari, and Dejan Stojkovic.
\newblock {Does DESI 2024 Confirm $\Lambda$CDM?}
\newblock 4 2024{\natexlab{b}}.

\bibitem[Colg\'ain et~al.(2024{\natexlab{c}})Colg\'ain, Pourojaghi, and
  Sheikh-Jabbari]{Colgain:2024ksa}
Eoin~\'O. Colg\'ain, Saeed Pourojaghi, and M.~M. Sheikh-Jabbari.
\newblock {Implications of DES 5YR SNe Dataset for $\Lambda$CDM}.
\newblock 6 2024{\natexlab{c}}.

\bibitem[Scolnic et~al.(2018)]{Pan-STARRS1:2017jku}
D.~M. Scolnic et~al.
\newblock {The Complete Light-curve Sample of Spectroscopically Confirmed SNe
  Ia from Pan-STARRS1 and Cosmological Constraints from the Combined Pantheon
  Sample}.
\newblock \emph{Astrophys. J.}, 859\penalty0 (2):\penalty0 101, 2018.
\newblock \doi{10.3847/1538-4357/aab9bb}.

\bibitem[Jimenez and Loeb(2002)]{Jimenez:2001gg}
Raul Jimenez and Abraham Loeb.
\newblock {Constraining cosmological parameters based on relative galaxy ages}.
\newblock \emph{Astrophys. J.}, 573:\penalty0 37--42, 2002.
\newblock \doi{10.1086/340549}.

\bibitem[Seo and Eisenstein(2003)]{Seo:2003pu}
Hee-Jong Seo and Daniel~J. Eisenstein.
\newblock {Probing dark energy with baryonic acoustic oscillations from future
  large galaxy redshift surveys}.
\newblock \emph{Astrophys. J.}, 598:\penalty0 720--740, 2003.
\newblock \doi{10.1086/379122}.

\bibitem[Eisenstein et~al.(2005{\natexlab{b}})]{SDSS:2005xqv}
Daniel~J. Eisenstein et~al.
\newblock {Detection of the Baryon Acoustic Peak in the Large-Scale Correlation
  Function of SDSS Luminous Red Galaxies}.
\newblock \emph{Astrophys. J.}, 633:\penalty0 560--574, 2005{\natexlab{b}}.
\newblock \doi{10.1086/466512}.

\bibitem[Lusso et~al.(2020)]{Lusso:2020pdb}
E.~Lusso et~al.
\newblock {Quasars as standard candles III. Validation of a new sample for
  cosmological studies}.
\newblock \emph{Astron. Astrophys.}, 642:\penalty0 A150, 2020.
\newblock \doi{10.1051/0004-6361/202038899}.

\bibitem[{Stern} et~al.(2010){Stern}, {Jimenez}, {Verde}, {Kamionkowski}, and
  {Stanford}]{2010JCAP...02..008S}
Daniel {Stern}, Raul {Jimenez}, Licia {Verde}, Marc {Kamionkowski}, and S.~Adam
  {Stanford}.
\newblock {Cosmic chronometers: constraining the equation of state of dark
  energy. I: H(z) measurements}.
\newblock \emph{\jcap}, 2010\penalty0 (2):\penalty0 008, February 2010.
\newblock \doi{10.1088/1475-7516/2010/02/008}.

\bibitem[Moresco et~al.(2012)]{2012JCAP...08..006M}
M.~Moresco et~al.
\newblock {Improved constraints on the expansion rate of the Universe up to z
  \raisebox{-0.5ex}\textasciitilde 1.1 from the spectroscopic evolution of
  cosmic chronometers}.
\newblock \emph{\jcap}, 2012\penalty0 (8):\penalty0 006, August 2012.
\newblock \doi{10.1088/1475-7516/2012/08/006}.

\bibitem[{Zhang} et~al.(2014){Zhang}, {Zhang}, {Yuan}, {Liu}, {Zhang}, and
  {Sun}]{2014RAA....14.1221Z}
Cong {Zhang}, Han {Zhang}, Shuo {Yuan}, Siqi {Liu}, Tong-Jie {Zhang}, and
  Yan-Chun {Sun}.
\newblock {Four new observational H(z) data from luminous red galaxies in the
  Sloan Digital Sky Survey data release seven}.
\newblock \emph{Research in Astronomy and Astrophysics}, 14\penalty0
  (10):\penalty0 1221-1233, October 2014.
\newblock \doi{10.1088/1674-4527/14/10/002}.

\bibitem[Moresco et~al.(2016)Moresco, Pozzetti, Cimatti, Jimenez, Maraston,
  Verde, Thomas, Citro, Tojeiro, and Wilkinson]{Moresco:2016mzx}
Michele Moresco, Lucia Pozzetti, Andrea Cimatti, Raul Jimenez, Claudia
  Maraston, Licia Verde, Daniel Thomas, Annalisa Citro, Rita Tojeiro, and David
  Wilkinson.
\newblock {A 6\% measurement of the Hubble parameter at $z\sim0.45$: direct
  evidence of the epoch of cosmic re-acceleration}.
\newblock \emph{JCAP}, 05:\penalty0 014, 2016.
\newblock \doi{10.1088/1475-7516/2016/05/014}.

\bibitem[Ratsimbazafy et~al.(2017)Ratsimbazafy, Loubser, Crawford, Cress,
  Bassett, Nichol, and V\"ais\"anen]{Ratsimbazafy:2017vga}
A.~L. Ratsimbazafy, S.~I. Loubser, S.~M. Crawford, C.~M. Cress, B.~A. Bassett,
  R.~C. Nichol, and P.~V\"ais\"anen.
\newblock {Age-dating Luminous Red Galaxies observed with the Southern African
  Large Telescope}.
\newblock \emph{Mon. Not. Roy. Astron. Soc.}, 467\penalty0 (3):\penalty0
  3239--3254, 2017.
\newblock \doi{10.1093/mnras/stx301}.

\bibitem[Borghi et~al.(2022)Borghi, Moresco, Cimatti, Huchet, Quai, and
  Pozzetti]{Borghi:2021zsr}
Nicola Borghi, Michele Moresco, Andrea Cimatti, Alexandre Huchet, Salvatore
  Quai, and Lucia Pozzetti.
\newblock {Toward a Better Understanding of Cosmic Chronometers: Stellar
  Population Properties of Passive Galaxies at Intermediate Redshift}.
\newblock \emph{Astrophys. J.}, 927\penalty0 (2):\penalty0 164, 2022.
\newblock \doi{10.3847/1538-4357/ac3240}.

\bibitem[Jiao et~al.(2023)Jiao, Borghi, Moresco, and Zhang]{Jiao:2022aep}
Kang Jiao, Nicola Borghi, Michele Moresco, and Tong-Jie Zhang.
\newblock {New Observational H(z) Data from Full-spectrum Fitting of Cosmic
  Chronometers in the LEGA-C Survey}.
\newblock \emph{Astrophys. J. Suppl.}, 265\penalty0 (2):\penalty0 48, 2023.
\newblock \doi{10.3847/1538-4365/acbc77}.

\bibitem[Tomasetti et~al.(2023)Tomasetti, Moresco, Borghi, Jiao, Cimatti,
  Pozzetti, Carnall, McLure, and Pentericci]{Tomasetti:2023kek}
E.~Tomasetti, M.~Moresco, N.~Borghi, K.~Jiao, A.~Cimatti, L.~Pozzetti, A.~C.
  Carnall, R.~J. McLure, and L.~Pentericci.
\newblock {A new measurement of the expansion history of the Universe at z =
  1.26 with cosmic chronometers in VANDELS}.
\newblock \emph{Astron. Astrophys.}, 679:\penalty0 A96, 2023.
\newblock \doi{10.1051/0004-6361/202346992}.

\end{thebibliography}
\label{lastpage}

\appendix
\setcounter{figure}{0} 
\renewcommand{\thefigure}{A\arabic{figure}} 
\setcounter{table}{0} 
\renewcommand{\thetable}{A\arabic{table}} 
\section{Polynomial fits and Cross Validation}\label{app:apx1}

\begin{figure*}
    \centering
    \includegraphics[width=9cm]{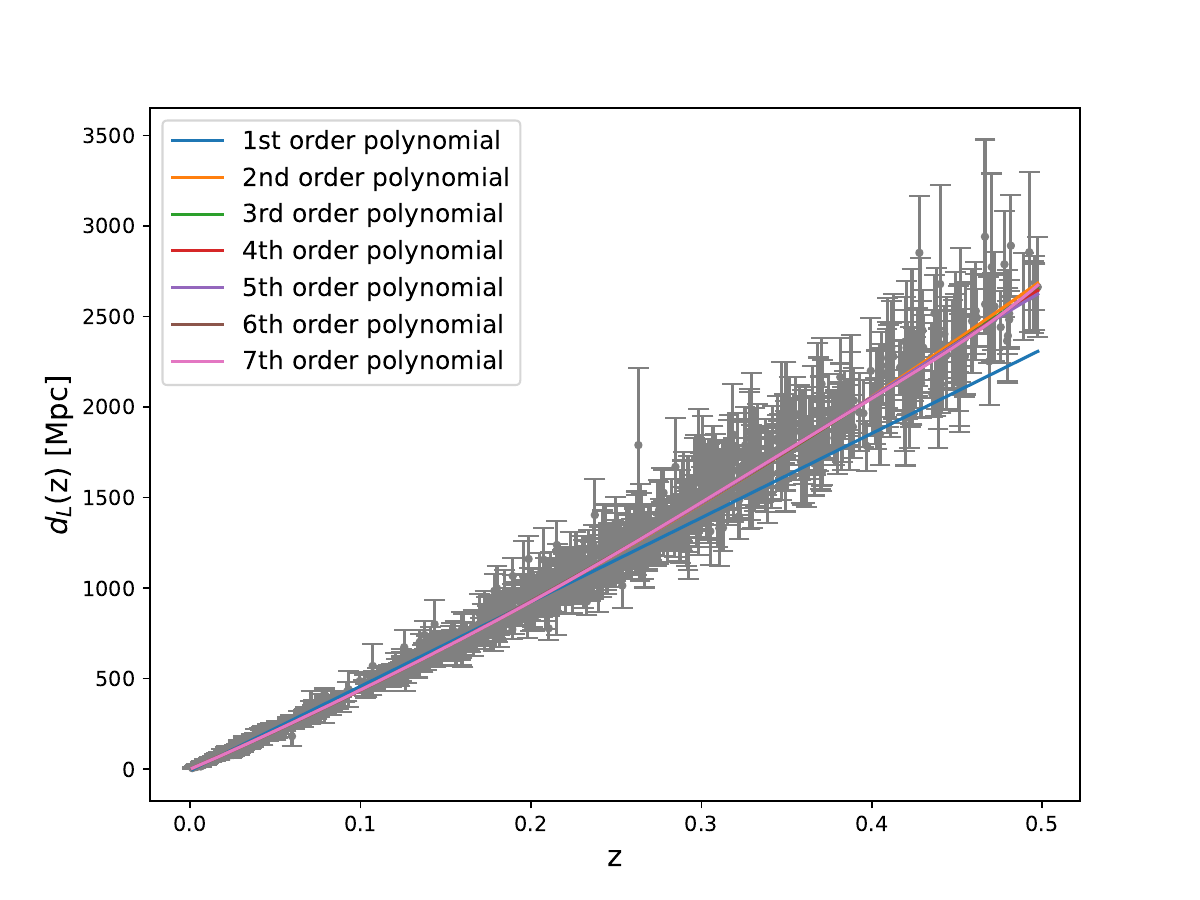}\includegraphics[width=9cm]{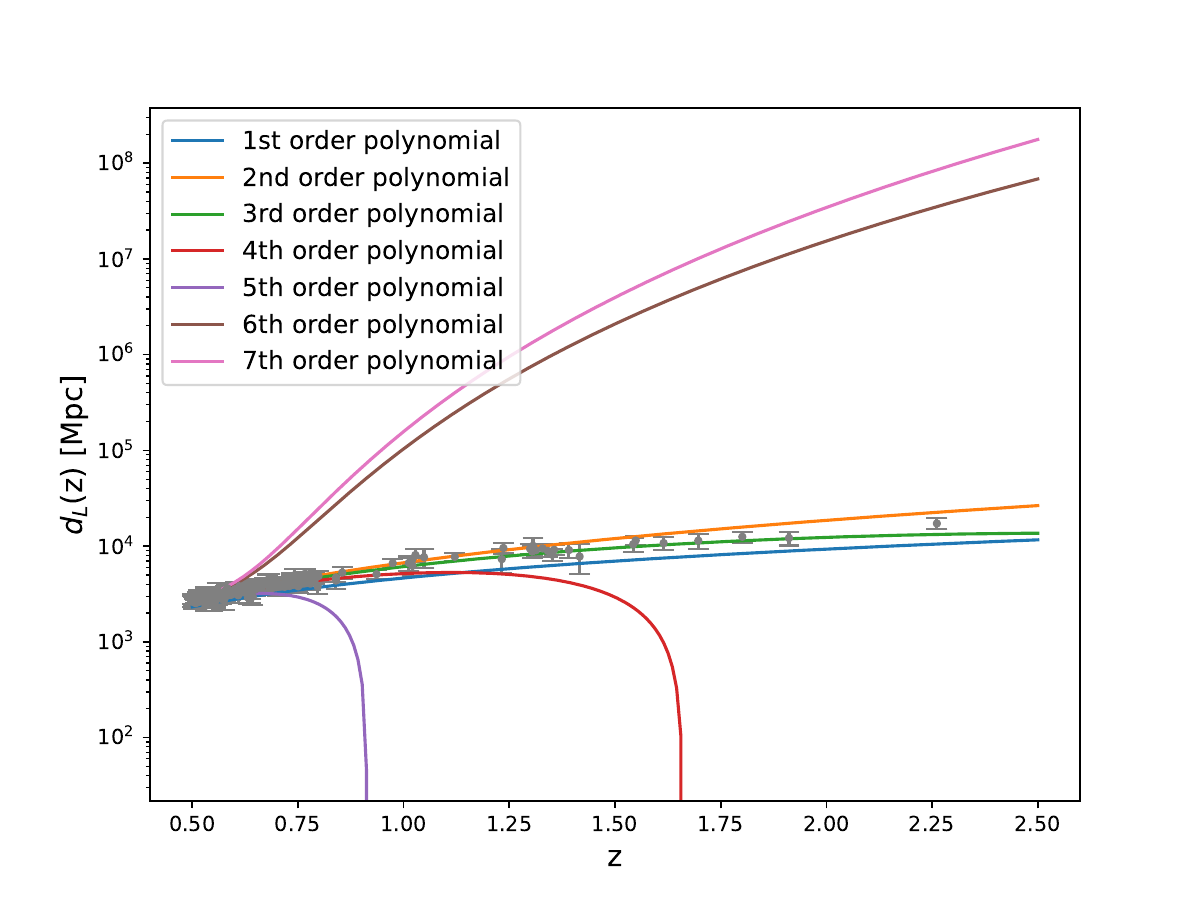}
    \caption{Fitting various polynomials to Pantheon+ data at $z < 1$ (left) and testing  them using Pantheon+ data at $z > 1$ (right).}
    \label{fig:fit_SN_low}
\end{figure*}

In this section, we describe the selection process for polynomials used in generating mock data. As outlined in Sec.~\ref{sect:mock}, a third-order polynomial was used to generate SNIa mock data, and a second-order polynomial was used for CC mock data. To identify the best polynomial fit, we apply the Cross-Validation technique as follows.
For both SNIa and CC data, various polynomials (see Eq.~\ref{eq:polynomial}) are fitted to approximately $80~\%$ of the all data points as training data. Notice that we choose  $80~\%$  of the observational data for SNIa are at $z < 0.5$ and for CC are at $z < 1$ as the training data.
We then determine the coefficients for each polynomial fit and calculate the corresponding $\chi^2$ values.
These polynomials, with coefficients determined from the train data, are then applied to the test data ($z > 0.5$ for SNIa and $z > 1$ for CC data).
We finally calculate the $\chi^2$ values for each polynomial using the test data to identify the best fit.
The results for Pantheon+ SNIa observations are presented in Table \ref{tab:fit_SN} and Figure \ref{fig:fit_SN_low}.
We observe that the third-order polynomial fit provide the best results for both training and test data. Moreover, fourth-order and higher-order polynomials failed to fit the test data ($z > 0.5$). Consequently, observational data at $z > 0.5$ cannot be encoded in mock data using fourth-order or higher-order polynomials.
Additionally, the first-order polynomial yielded $\chi^2 = 1237.02$ for $z < 0.5$ and $\chi^2 = 570.61$ for $z > 0.5$, indicating that it does not fit the training data well and eventually differs significantly from the test data. Among the different polynomials, the third-order polynomial provides the best fit to the training data and also has the minimum $\chi^2$ value for the test data. Therefore, we select it as the best polynomial fit for generating mock SNIa data.
\begin{align}\label{eq:polynomial}
y_1 &= a + bz\;, \nonumber \\
y_2 &= a + bz + cz^2\;, \nonumber \\
y_3 &= a + bz + cz^2 + dz^3\;, \nonumber \\
y_4 &= a + bz + cz^2 + dz^3 + ez^4\;, \nonumber \\
y_5 &= a + bz + cz^2 + dz^3 + ez^4 + fz^5\;, \nonumber \\
y_6 &= a + bz + cz^2 + dz^3 + ez^4 + fz^5 + gz^6\;, \nonumber \\
y_7 &= a + bz + cz^2 + dz^3 + ez^4 + fz^5 + gz^6 + hz^7\;.
\end{align}

\begin{figure*}
    \centering
    \includegraphics[width=9cm]{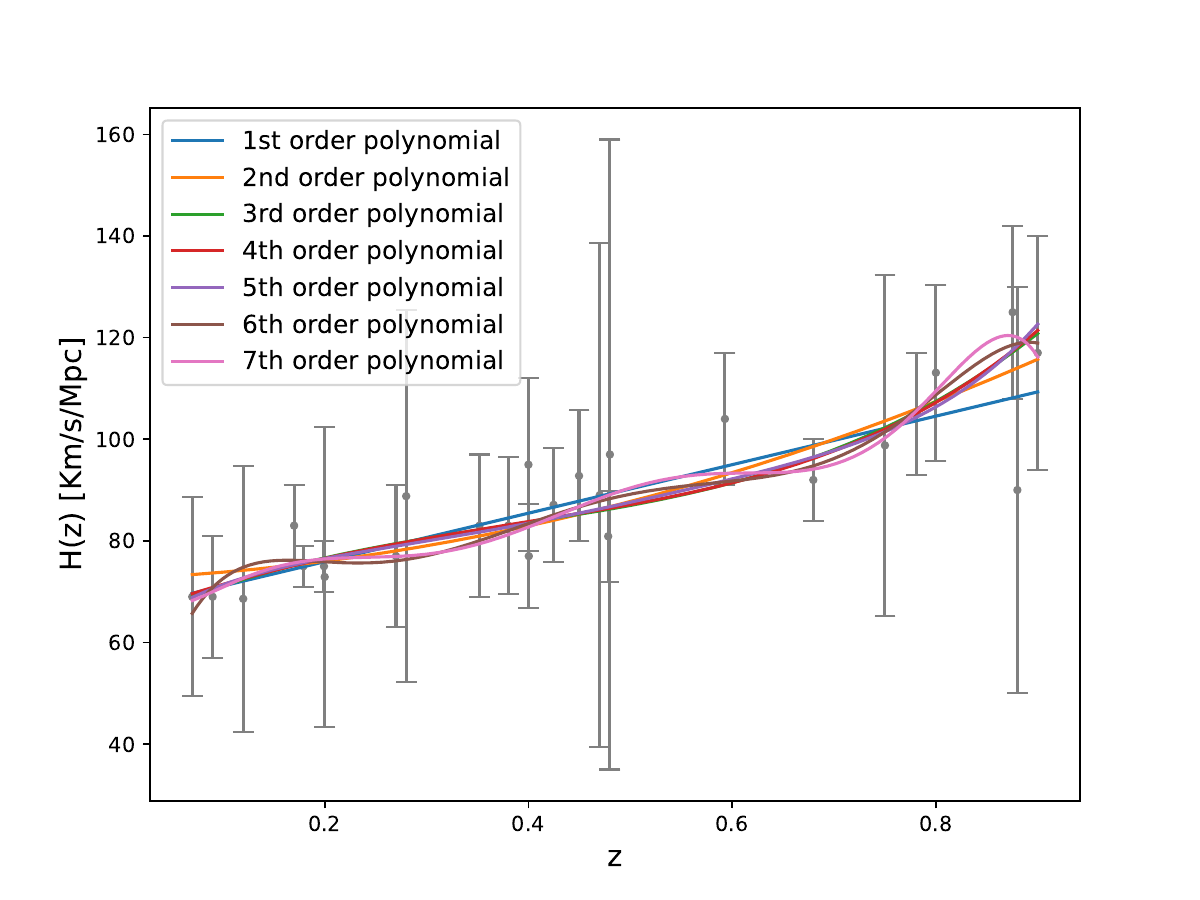}\includegraphics[width=9cm]{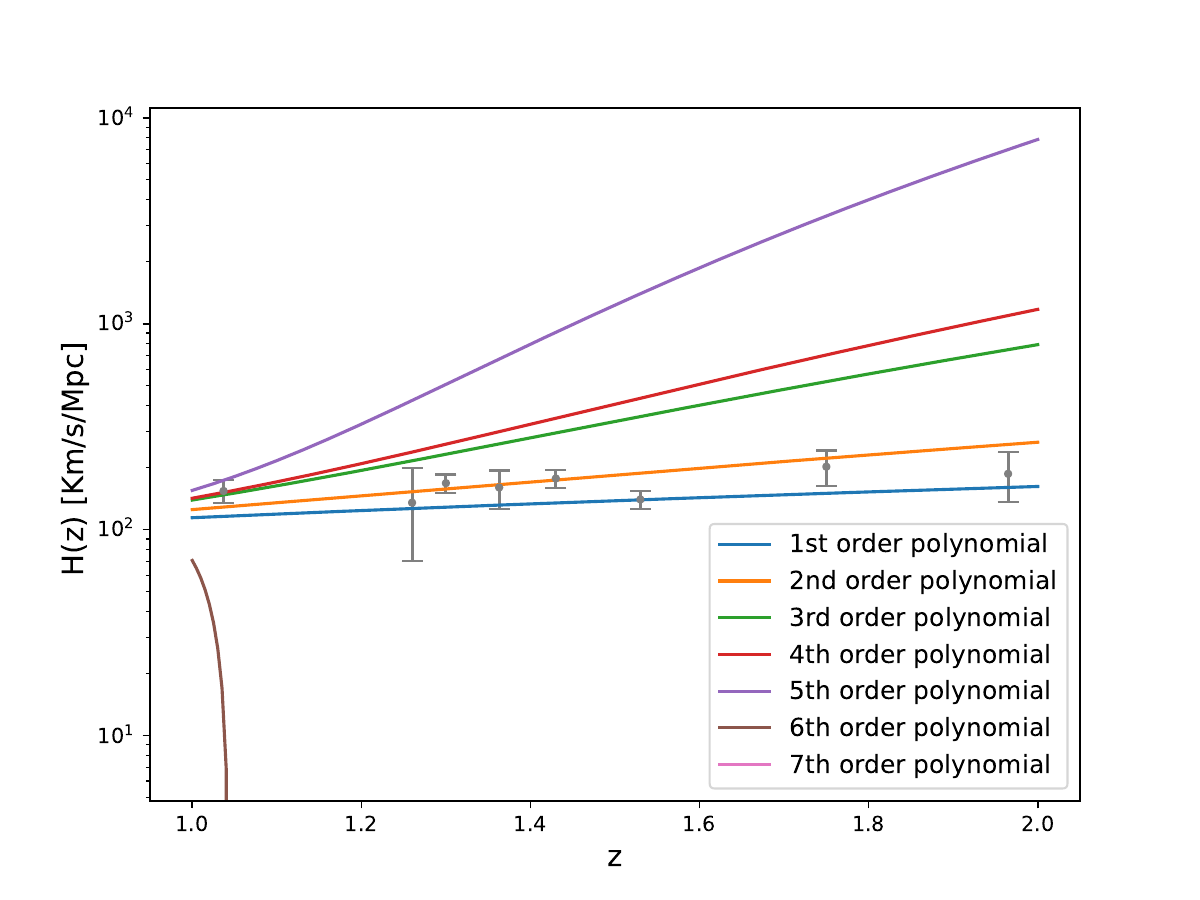}
    \caption{Fitting polynomials to CC train data ($z < 1$) (left) and examine them using test data($z > 1$) (right).}
    \label{fig:fit_CC_low}
\end{figure*}

\begin{table}
\centering
\caption{The $\chi^2$ values for different polynomials trained using Pantheon+ data at $z < 0.5$, and tested using Pantheon+ data at $z > 0.5$.}
\begin{tabular}{c c c}
\hline \hline
Polynomial Order & $\chi^2$ for low-$z$ data & $\chi^2$ for high-$z$.\\
\hline 
1st order & $1273.02$ & $570.61$ \\
2nd order & $677.97$ & $123.73$ \\
3rd order & $676.45$ & $75.62$ \\
4th order & $676.35$ & $993.81$ \\
5th order & $676.07$ & $422526.73$ \\
6th order & $674.79$ & $323200571.81$ \\
7th order & $674.78$ & $1781368299.93$ \\
\hline \hline
\end{tabular}
\label{tab:fit_SN}
\end{table}
For CC data, we follow the same procedure as for the SNIa Pantheon+ data to determine the best polynomial fit to the observations. The results are presented in Figure~\ref{fig:fit_CC_low} and Table~\ref{tab:fit_cc}. We observe that, due to the large uncertainties in CC data, all polynomials fit the data at $z < 1$ well. The $\chi^2$ values for fitting different polynomials to the training data ($z < 1$) are similar, with negligible differences. However, for the test data ($z > 1$), only the first- and second-order polynomials align with the observational data (see also the right panel of Figure~\ref{fig:fit_CC_low}). Consequently, we select the second-order polynomial as the best fit and use it to generate mock CC data. Finally, we emphasize that for both the SNIa and CC analyses, we divided the data into five folds for training and testing. In each fold, we used $80\%$ of the total data for training and the remaining $20\%$ for testing. Consequently, after analyzing all folds, we found that a second-order polynomial is the best fit for the SNIa observations, while a third-order polynomial is the best fit for the CC observations.

\begin{table}
\centering
\caption{The $\chi^2$ values for different polynomials trained by CC data ($z < 1$) and examined by test data ($z > 1$).}
\begin{tabular}{c c c}
\hline \hline
 Polynomial Order & $\chi^2$ for low-$z$ data & $\chi^2$ for high-$z$ data. \\
\hline
1st order & $6.15$ & $17.30$ \\
2nd order & $5.40$ & $16.39$ \\
3rd order & $5.02$ & $489.50$ \\
4th order & $5.01$ & $1067.30$ \\
5th order & $4.99$ & $34883.90$ \\
6th order & $4.70$ & $10892303.00$ \\
7th order & $4.57$ & $392381814.90$ \\
\hline \hline
\end{tabular}
\label{tab:fit_cc}
\end{table}

\setcounter{figure}{0} 
\renewcommand{\thefigure}{B\arabic{figure}} 
\section{Testing our generating mock data mechanism}\label{app:apx2}
\begin{figure}
    \centering
    \includegraphics[width=8cm]{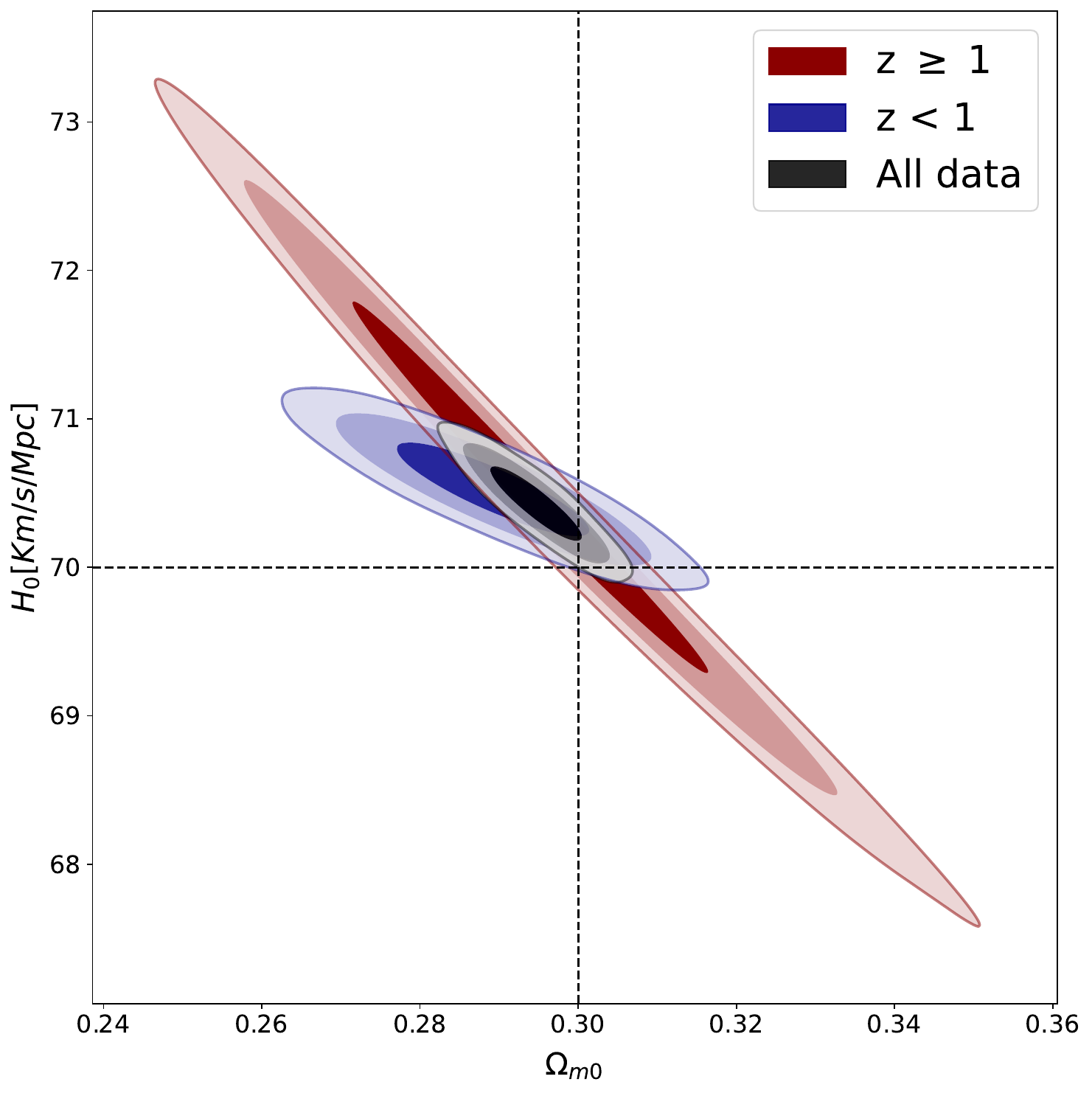}
    \caption{Confidence regions for \(\Omega_{m0}\) and \(H_0\), using 10,000 mock SNIa data points generated from the Planck-\(\Lambda\)CDM model. Results shown for all data, high-\(z\), and low-\(z\) datasets are recovering the canonical values $H_0=70$km/s/Mpc \& $\Omega_{m0}=0.3$ within $3\sigma$ confidence regions.}
    \label{fig:corner_LCDM}
\end{figure}

In order to test our method for generating mock data in Sec.~\ref{sect:mock} and consequently validate our mock analysis, we generate mock data based on the Planck-$\Lambda$CDM model. In this case, we expect to observe no deviation between our constraints on the cosmological parameters from high-redshift and low-redshift mock data. To verify this, we generate $N=10000$ mock distance modulus \(\mu(z_i)\) using Eq.~\ref{eq7}, with a redshift range of \(0 < z < 2.5\). The input parameter values are \(\Omega_{m0} = 0.3\) and \(H_0 = 70 km/s/Mpc\). Our results for the cosmological constraints on the parameters $\Omega_{m0}$ and $H_0$ using high-redshift (\(z \geq 1\)) and low-redshift (\(z < 1\)) datasets are shown in Fig.~\ref{fig:corner_LCDM}. We observe that there is no significant deviation in the values of the cosmological parameters obtained from the high-redshift and low-redshift data. Both results recover the input parameters within the \(3\sigma\) confidence level.
In contrast, we observed that when mock data are generated using a polynomial fit to observations, significant deviations arise within the $\Lambda$CDM model fitting to low-redshift and high-redshift datasets. This suggests that if the real observational data follow the $\Lambda$CDM model, we should expect to observe no significant deviation between the cosmological constraints obtained from low-redshift and high-redshift observations. An expectation that we did not observe in our mock analysis.

\end{document}